\documentclass[aps,prb,twocolumn,showpacs]{revtex4}

\usepackage{graphicx}
\usepackage{amsmath}
\usepackage{amssymb}

\begin{document}

\title{Influence of impurities on electronic structure in cuprate superconductors}

\author{Minghuan Zeng\footnotemark[1], Xiang Li\footnotemark[1]\footnotetext[1]{These authors
contributed equally to this work}, Yongjun Wang, and Shiping Feng\footnote[2]{E-mail address:
spfeng@bnu.edu.cn}}

\affiliation{Department of Physics, Beijing Normal University, Beijing 100875, China}

%\date{\today}

\begin{abstract}
The impurity is inherently manifest in cuprate superconductors, as cation substitution or
intercalation is necessary for the introduction of charge carriers, and its influence on
the electronic state is at the heart of a great debate in physics. Here based on the
{\it microscopic octet scattering model}, the influence of the impurity scattering on the
electronic structure of cuprate superconductors is investigated in terms of the
self-consistent $T$-matrix approach. The impurity scattering self-energy is evaluated
firstly in the {\it Fermi-arc-tip approximation} of the quasiparticle excitations and
scattering processes, and the obtained results show that the decisive role played by the
impurity scattering self-energy in the particle-hole channel is the further renormalization
of the quasiparticle band structure with a reduced quasiparticle lifetime, while the
impurity scattering self-energy in the particle-particle channel induces a strong deviation
from the d-wave behaviour of the superconducting gap, leading to the existence of a finite
gap over the entire electron Fermi surface. Moreover, these impurity scattering self-energies
are employed to study the exotic features of the line-shape in the quasiparticle excitation
spectrum and the autocorrelation of the quasiparticle excitation spectra, and the obtained
results are then compared with the corresponding experimental data. The theory therefore also
indicates that the unconventional features of the electronic structure in cuprate
superconductors is generated by both the strong electron correlation and impurity scattering.
\end{abstract}


\maketitle

\section{INTRODUCTION}\label{Introduction}

The single common feature in the crystal structure of cuprate superconductors is the presence
of the square-lattice CuO$_{2}$ layer \cite{Bednorz86,Wu87}, which are believed to contain
all the essential physics. The layered crystal structure then is a stacking of CuO$_{2}$
layers separated by other oxide layers, which maintain the charge neutrality and cohesion of
the structure mainly through ionic interactions \cite{Bednorz86,Wu87}. The parent compound of
cuprate superconductors is a Mott insulator with an antiferromagnetic (AF) long-range order
\cite{Fujita12}, and superconductivity then is realized when this AF long-range order state
is suppressed by doped charge carriers into the CuO$_{2}$ layer \cite{Bednorz86,Wu87}. In
addition to the change of the charge-carrier concentration, this doping process nearly always
introduces some measure of disorder \cite{Hussey02,Balatsky06,Alloul09}, leading to that in
principle, all cuprate superconductors have naturally impurities (or disorder). In particular,
impurities which substitute for Cu in the CuO$_{2}$ layer turn out to be strong scatters of
the electronic state in the layer \cite{Hussey02,Balatsky06,Alloul09}. The importance of the
understanding of the influence of the impurity scattering on the electronic structure has been
quickly recognized, since many of the unconventional features, including the relatively high
superconducting (SC) transition temperature $T_{\rm c}$, have often been attributed to
particular characteristics of the low-energy quasiparticle excitations determined by the
electronic structure \cite{Damascelli03,Campuzano04,Fink07}.

The impurity scattering in cuprate superconductors is specially unconventional, and manifests
a variety of the phenomena depending on the strength of the electron correlation,
charge-carrier doping, temperature, and magnetic field \cite{Hussey02,Balatsky06,Alloul09}.
Experimentally, by virtue of systematic studies using multiple measurement techniques, a
number of consequences from the impurity scattering together with the associated fluctuation
phenomena have been identified \cite{Hussey02,Balatsky06,Alloul09}, where an agreement has
emerged that the various properties of the d-wave SC-state in the pure system are extreme
sensitivity to the influence of the impurity scattering than that in the conventional
superconductors. This follows a basic fact that the influence of the impurity scattering on
the d-wave SC-state is to break the electron pairs and to mix the SC gap with different signs
on different parts of the electron Fermi surface (EFS) \cite{Hussey02,Balatsky06,Alloul09}.
In particular, the early experimental measurements
\cite{Ishida91,Legris93,Giapintzakis94,Fukuzumi96,Attfield98,Bobroff99,Eisaki04} showed that
the influence of the impurity scattering tends to suppress the SC coherence, and then
$T_{\rm c}$ is found to be depressed rapidly with the increase of the impurity concentration.
Later, the experimental observations indicated that the extent of the deviation from the
d-wave SC gap form increases with the increase of the impurity concentration
\cite{Vobornik99,Vobornik00,Shen04,Kondo07,Pan09}. This impurity concentration dependence of
the suppression of $T_{\rm c}$ thus has been reflected in the nature of the SC-state
quasiparticle excitations resulting of the dressing of the electrons via the impurity
scattering, where a change from linear temperature dependence to the quadratic behavior in
the magnetic-field penetration-depth occurs due to the influence of the impurity scattering
\cite{Bonn94}, while the ratio of the low-temperature superfluid density and the effective
mass of the electrons $n_{\rm s}(T\rightarrow 0)/m^{*}$ is observed experimentally to
decrease with the increase of the impurity concentration \cite{Bucci94,Bernhard96,Bobroff05}.
In particular, the angle-resolved photoemission spectroscopy (ARPES) experiments
\cite{Vobornik99,Vobornik00,Shen04,Kondo07,Pan09} indicate that the spectral linewidth of
the quasiparticle excitation spectrum broadens rapidly with the increase of the impurity
concentration, leading to that the spectral intensity is suppressed almost linearly in energy
at low temperatures. These experimental results therefore offer experimental evidences that
the electronic structure and SC-state properties in the pure cuprate superconductors are
significantly influenced by the impurity scattering.

Although a number of consequences from the impurity scattering
\cite{Hussey02,Balatsky06,Alloul09} together with the associated fluctuation phenomena have
been well-identified experimentally
\cite{Vobornik99,Vobornik00,Shen04,Kondo07,Pan09,Bonn94,Bucci94,Bernhard96,Bobroff05}, the
full understanding of the influence of the impurity scattering on the electronic state is
still a challenging issue. Theoretically, the homogenous part of the SC-state electron
propagator in the preceding discussions is based on the modified Bardeen-Cooper-Schrieffer
(BCS) formalism with the d-wave symmetry \cite{Hussey02,Balatsky06,Alloul09}, and then the
coupling between the electrons and impurities as the perturbation is treated in terms of the
self-consistent T-matrix approach for a single impurity or a finite impurity concentration
\cite{Hussey02,Balatsky06,Alloul09,Mahan81,Hirschfeld89,Hirschfeld93}. In particular, the
characteristic feature of the d-wave SC-state is the existence of four nodes on EFS, where
the SC gap vanishes, and then the SC-state properties are largely governed by the
quasiparticle excitations at around the nodal region \cite{Hussey02,Balatsky06,Alloul09}. In
this case, the impurity scattering self-energy was evaluated in the nodal approximation of
the quasiparticle excitations and scattering processes
\cite{Hussey02,Balatsky06,Alloul09}, and was used to discuss the various properties of the
SC-state in cuprate superconductors
\cite{Hussey02,Balatsky06,Alloul09,Durst00,Yashenkin01,Nunner05,Dahm05,Wang08,Andersen08,Wang09,Hone17}.
However, it has been demonstrated experimentally \cite{Chatterjee06,He14} that the Fermi
arcs formed by the disconnected segments on the constant energy contour that emerge due to
the EFS reconstruction at the case of zero energy \cite{Shi08,Sassa11,Comin14,Horio16,Loret18}
can persist into the case for a finite binding-energy, where the quasiparticle scattering
further reduces almost all spectral weight on Fermi arcs to the tips of the Fermi arcs, and
then the most physical properties are mainly controlled by the quasiparticle excitations at
around the tips of the Fermi arcs. Moreover, these tips of the Fermi arcs connected by the
scattering wave vectors ${\bf q}_{i}$ construct a {\it octet scattering model}, and then the
quasiparticle scattering processes with the scattering wave vectors ${\bf q}_{i}$ contribute
effectively to the quasiparticle scattering processes \cite{Chatterjee06,He14}. It should be
emphasized that this octet scattering model is a basic model in the explanation of the
Fourier transform scanning tunneling spectroscopy experimental data
\cite{Yin21,Pan01,Hoffman02,Kohsaka07,Kohsaka08,Hamidian16}, and also can give a consistent
description of the regions of the highest joint density of states detected from the ARPES
autocorrelation experiments \cite{Chatterjee06,He14}. However, to the best of our knowledge,
the influence of the impurity scattering on the electronic structure has not been discussed
starting from the {\it microscopic octet scattering model} to treat the impurity scattering
in terms of the self-consistent $T$-matrix approach, and no explicit calculation of the
impurity scattering self-energy has been made so far in the {\it Fermi-arc-tip approximation}
of the quasiparticle excitations and scattering processes.

In this paper, we start from the homogenous part of the electron propagator and the related
{\it microscopic octet scattering model} obtained within the framework of the
kinetic-energy-driven superconductivity \cite{Feng0306,Feng12,Feng15a,Feng15} to study the
influence of the impurity scattering on the electronic structure of cuprate superconductors
in terms of the self-consistent $T$-matrix approach, where we evaluate firstly the impurity
scattering self-energy in the {\it Fermi-arc-tip approximation} of the quasiparticle
excitations and scattering processes, and the obtained results show that (i) the crucial
role of the impurity scattering self-energy in the particle-hole channel is the further
renormalization of the quasiparticle band structure and reduction of the quasiparticle
lifetime with the renormalization strength that increase as the impurity concentration is
increased; (ii) the impurity scattering self-energy in the particle-particle channel induces
a strong deviation from the d-wave behaviour of the SC gap, leading to the existence of a
finite gap over the entire EFS. In particular, with the increase of the impurity
concentration, the magnitude of the SC gap is progressively decreased by the impurity
scattering self-energy along EFS except for at around the nodal region, where the magnitude
of the gap smoothly increases. Moreover, these impurity scattering self-energies are
employed to study the unconventional features of the line-shape in the quasiparticle
excitation spectrum and the ARPES autocorrelation spectrum, and the obtained results are
well consistent with the corresponding experimental data.

The rest of this paper is organized as follows. We derive explicitly the dressed electron
propagator in Sec. \ref{Formalism}, and then employ this dressed electron propagator to
discuss the impurity dependence of the electronic structure in
Sec. \ref{Quantitative-characteristics}, where we show that in addition to the suppression
of the spectral weight in the quasiparticle excitation spectrum, the position of the
low-energy coherent peak at around the antinodal region is shifted towards to EFS when the
impurity concentration is increased, while the position of the low-energy coherent peak at
around the nodal region moves away from EFS. In particular, the sharp peaks in the ARPES
autocorrelation spectrum are directly correlated with the scattering wave vectors
${\bf q}_{i}$ connecting the tips of the Fermi arcs, and then the key signature of the
Fermi-arc-tip quasiparticle correlation appears in the ARPES autocorrelation spectrum, which
is essentially quasiparticle scattering interference. Finally, we give a summary and
discussions in Sec. \ref{conclude}. In the Appendix, we presents the details of the
derivation of the dressed electron propagator.

\section{Formalism}\label{Formalism}

\subsection{$t$-$J$ model and homogenous electron propagator}
\label{model-propagator}

To set the stage for the discussion of the influence of the impurity scattering on the
electronic structure of cuprate superconductors, we first give an account of the model and
homogenous electron propagator used to describe the intrinsic aspects of the pure cuprate
superconductors. As we have mentioned above, all the essential important in cuprate
superconductors are contained in the doped CuO$_{2}$ layer. Shortly after the discovery of
superconductivity in cuprate superconductors \cite{Bednorz86}, it was proposed that the
$t$-$J$ model on a square lattice is an appropriate model to describe the essential physics
of the doped CuO$_{2}$ layer \cite{Anderson87},
\begin{eqnarray}\label{tjham}
H=-\sum_{\langle l\hat{a}\rangle\sigma}t_{l\hat{a}}C^{\dagger}_{l\sigma}C_{l+\hat{a}\sigma}
+\mu\sum_{l\sigma}C^{\dagger}_{l\sigma}C_{l\sigma}+J\sum_{\langle l\hat{\eta}\rangle}
{\bf S}_{l}\cdot {\bf S}_{l+\hat{\eta}},~
\end{eqnarray}
where the double electron occupancy is no longer allowed, i.e.,
$\sum_{\sigma}C^{\dagger}_{l\sigma}C_{l\sigma}\leq 1$, $C^{\dagger}_{l\sigma}$ and
$C_{l\sigma}$ are creation and annihilation operators for the constrained electrons with spin
orientation $\sigma=\uparrow,\downarrow$ on lattice site $l$, respectively,
${\bf S}_{l}=(S^{\rm x}_{l},S^{\rm y}_{l},S^{\rm z}_{l})$ is spin operator, $\mu$ is the
chemical potential, and $J$ is the exchange coupling between the nearest-neighbor (NN) sites
$\hat{\eta}$. In this paper, the hopping of the constrained electrons $t_{l\hat{a}}$ is
restricted to the NN sites $\hat{\eta}$ and next NN sites $\hat{\tau}$ with the amplitudes
$t_{l\hat{\eta}}=t$ and $t_{l\hat{\tau}}=-t'$, respectively, while the summation
$\langle l\hat{a}\rangle$ denotes that $l$ runs over all sites, and for each $l$, over its NN
sites $\hat{a}=\hat{\eta}$ or next NN sites $\hat{a}=\hat{\tau}$. Hereafter, the parameters
are chosen as $t/J=2.5$ and $t'/t=0.3$ as in the previous discussions \cite{Feng15a,Feng15}.
The magnitude of $J$ and the lattice constant of the square lattice are the energy and length
units, respectively. However, when necessary to compare with the experimental data, we set
$J=100$ meV. The strong electron correlation in cuprate superconductors manifests itself by
the on-site local constraint of no double electron occupancy, and this is why the crucial
requirement is to impose this on-site local constraint properly
\cite{Yu92,Feng93,Zhang93,Guillou95,Lee06}. In particular, it has been shown that this
on-site local constraint can be fulfilled in the fermion-spin approach
\cite{Feng15,Feng0494}, where the constrained electron operators $C_{l\uparrow}$ and
$C_{l\downarrow}$ in the $t$-$J$ model (\ref{tjham}) are replaced by,
\begin{eqnarray}\label{css}
C_{l\uparrow}=h^{\dagger}_{l\uparrow}S^{-}_{l},~~~~
C_{l\downarrow}=h^{\dagger}_{l\downarrow}S^{+}_{l},
\end{eqnarray}
with the spinful fermion operator $h_{l\sigma}=e^{-i\Phi_{l\sigma}}h_{l}$ that represents
the charge degree of freedom of the constrained electron together with some effects of spin
configuration rearrangements due to the presence of the doped hole itself (charge carrier),
while the spin operator $S_{l}$ describes the spin degree of freedom of the constrained
electron, and then the local constraint of no double occupancy at each site is fulfilled in
actual analyses. In this fermion-spin representation, the original $t$-$J$ model
(\ref{tjham}) can be rewritten explicitly as,
\begin{eqnarray}\label{cssham}
H&=&\sum_{\langle l\hat{a}\rangle}t_{l\hat{a}}(h^{\dagger}_{l+\hat{a}\uparrow}h_{l\uparrow}
S^{+}_{l}S^{-}_{l+\hat{a}} +h^{\dagger}_{l+\hat{a}\downarrow}h_{l\downarrow}S^{-}_{l}
S^{+}_{l+\hat{a}}) \nonumber\\
&-&\mu_{\rm h}\sum_{l\sigma}h^{\dagger}_{l\sigma}h_{l\sigma}+J_{{\rm eff}}
\sum_{\langle l\hat{\eta}\rangle}{\bf S}_{l}\cdot {\bf S}_{l+\hat{\eta}},~~~~
\end{eqnarray}
where $S^{-}_{l}=S^{\rm x}_{l}-iS^{\rm y}_{l}$ and $S^{+}_{l}=S^{\rm x}_{l}+iS^{\rm y}_{l}$
are the spin-lowering and spin-raising operators for the spin $S=1/2$, respectively,
$J_{{\rm eff}}=(1-\delta)^{2}J$,
$\delta=\langle h^{\dagger}_{l\sigma}h_{l\sigma}\rangle=\langle h^{\dagger}_{l}h_{l}\rangle$
is the charge-carrier doping concentration, and $\mu_{\rm h}$ is the charge-carrier chemical
potential. Concomitantly, the kinetic-energy term in the $t$-$J$ model (\ref{tjham}) has been
transferred as the coupling between charge and spin degrees of freedom of the constrained
electron, and therefore dominates the essential physics in the pure cuprate superconductors.

For a microscopic description of the SC-state in the pure cuprate superconductors, the
kinetic-energy-driven SC mechanism has been established based on the $t$-$J$ model
(\ref{cssham}) in the fermion-spin representation \cite{Feng0306,Feng12,Feng15a,Feng15},
where the coupling between charge and spin degrees of freedom of the constrained electron
directly from the kinetic energy by the exchange of a strongly dispersive spin excitation
generates a d-wave charge-carrier pairing in the particle-particle channel, then the d-wave
electron pairs originated from the d-wave charge-carrier pairing state are due to the
charge-spin recombination \cite{Feng15a}, and their condensation reveals the d-wave SC-state.
The typical features of the kinetic-energy-driven SC mechanism can be summarized as
\cite{Feng0306,Feng12,Feng15a,Feng15}: (i) the mechanism is purely electronic without
phonons; (ii) the mechanism indicates that the strong electron correlation favors
superconductivity, since the main ingredient is identified into an electron pairing mechanism
not involving the phonon, the external degree of freedom, but the internal spin degree of
freedom of the constrained electron; (iii) the SC-state is controlled by both the SC gap and
quasiparticle coherence, leading to that the maximal $T_{\rm c}$ occurs around the optimal
doping, and then decreases in both the underdoped and the overdoped regimes. Following these
previous discussions, the homogenous electron propagator of the $t$-$J$ model (\ref{cssham})
in the SC-state can be expressed explicitly in the Nambu representation as \cite{Feng15a},
%\begin{widetext}
\begin{eqnarray}\label{EGF}
\tilde{G}({\bf k},\omega)&=&\left(
\begin{array}{cc}
G({\bf k},\omega), & \Im({\bf k},\omega) \\
\Im^{\dagger}({\bf k},\omega), & -G({\bf k},-\omega)
\end{array}\right)\nonumber\\
&=&{1\over F({\bf k},\omega)}\{[\omega-\Sigma_{0}({\bf k},\omega)]\tau_{0}
+\Sigma_{1}({\bf k},\omega)\tau_{1}\nonumber\\
&+&\Sigma_{2}({\bf k},\omega)\tau_{2}+[\varepsilon_{\bf k}
+\Sigma_{3}({\bf k},\omega)]\tau_{3}\},~~~
\end{eqnarray}
%\end{widetext}
where $\tau_{0}$ is the unit matrix, $\tau_{1}$, $\tau_{2}$, and $\tau_{3}$ are Pauli
matrices, $\varepsilon_{\bf k}=-4t\gamma_{\bf k}+4t'\gamma_{\bf k}'+\mu$ is the energy
dispersion in the tight-binding approximation, with
$\gamma_{\bf k}=({\rm cos}k_{x}+{\rm cos} k_{y})/2$,
$\gamma_{\bf k}'={\rm cos}k_{x}{\rm cos}k_{y}$,
$F({\bf k},\omega)=[\omega-\Sigma_{0}({\bf k},\omega)]^{2}-[\varepsilon_{\bf k}
+\Sigma_{3}({\bf k},\omega)]^{2}-\Sigma^{2}_{1}({\bf k},\omega)
-\Sigma^{2}_{2}({\bf k},\omega)$. The homogenous self-energy
$\Sigma_{\rm pp} ({\bf k},\omega)$ in the particle-particle channel is identified as the
energy and momentum dependence of the d-wave SC gap \cite{Eliashberg60},
while the homogenous self-energy $\Sigma_{\rm ph}({\bf k},\omega)$ in the particle-hole
channel represents the quasiparticle coherence. In particular,
$\Sigma_{\rm pp} ({\bf k},\omega)$ is an even function of $\omega$, while
$\Sigma_{\rm ph}({\bf k},\omega)$ is not. However, in the above expression of the homogenous
electron propagator (\ref{EGF}), $\Sigma_{\rm pp} ({\bf k},\omega)$ has been separated into
its real and imaginary parts as:
$\Sigma_{\rm pp}({\bf k},\omega)=\Sigma_{1}({\bf k},\omega)-i\Sigma_{2}({\bf k},\omega)$,
while $\Sigma_{\rm ph}({\bf k},\omega)$ has been broken up into its symmetric and
antisymmetric parts as:
$\Sigma_{\rm ph}({\bf k},\omega)=\Sigma_{3}({\bf k},\omega)+\Sigma_{0}({\bf k},\omega)$, and
then both $\Sigma_{0}({\bf k},\omega)/\omega$ and $\Sigma_{3}({\bf k},\omega)$ are an even
function of $\omega$. In this case, the components of the homogenous self-energy in the
particle-hole channel $\Sigma_{0}({\bf k},\omega)$ and $\Sigma_{3}({\bf k},\omega)$ satisfy
the following identities,
\begin{subequations}
\begin{eqnarray}
{\rm Re}\Sigma_{0}({\bf k},\omega) &=& -{\rm Re}\Sigma_{0}({\bf k},-\omega), \\
{\rm Im}\Sigma_{0}({\bf k},\omega) &=& {\rm Im}\Sigma_{0}({\bf k},-\omega), \\
{\rm Re}\Sigma_{3}({\bf k},\omega) &=& {\rm Re}\Sigma_{3}({\bf k},-\omega), \\
{\rm Im}\Sigma_{3}({\bf k},\omega) &=& -{\rm Im}\Sigma_{3}({\bf k},-\omega).
\end{eqnarray}
\end{subequations}

In the framework of the kinetic-energy-driven superconductivity
\cite{Feng0306,Feng12,Feng15a,Feng15}, both $\Sigma_{\rm ph}({\bf k},\omega)$ [then
$\Sigma_{0}({\bf k},\omega)$ and $\Sigma_{3}({\bf k},\omega)$] and
$\Sigma_{\rm pp}({\bf k},\omega)$ [then $\Sigma_{1}({\bf k},\omega)$ and
$\Sigma_{2}({\bf k},\omega)$] arise from the interaction between electrons mediated by a
strongly dispersive spin excitation, and have been derived explicitly in
Ref. \onlinecite{Feng15a} in terms of the full charge-spin recombination, where all order
parameters and chemical potential are determined by the self-consistent calculation without
using any adjustable parameters. In particular, the sharp peak visible for temperature
$T\rightarrow 0$ in $\Sigma_{\rm ph}({\bf k},\omega)$ [$\Sigma_{\rm pp}({\bf k},\omega)$] is
actually a $\delta$-functions, broadened by a small damping used in the numerical calculation
for a finite lattice. The calculation in this paper for $\Sigma_{\rm ph}({\bf k},\omega)$ and
$\Sigma_{\rm pp}({\bf k},\omega)$ is performed numerically on a $120\times 120$ lattice in
momentum space, with the infinitesimal $i0_{+}\rightarrow i\Gamma$ replaced by a small
damping $\Gamma=0.05J$.

\subsection{Octet scattering model}\label{Octet-model}

With the above homogenous electron propagator (\ref{EGF}), the homogenous electron spectral
function $A({\bf k},\omega)$ now can be obtained explicitly as,
\begin{eqnarray}\label{ESF}
A({\bf k},\omega)&=&-2{\rm Im}G({\bf k},\omega)\nonumber\\
&=&{-2{\rm Im}\Sigma_{\rm tot}({\bf k},\omega)\over [\omega-\varepsilon_{\bf k}
-{\rm Re}\Sigma_{\rm tot}({\bf k},\omega)]^{2}
+[{\rm Im}\Sigma_{\rm tot}({\bf k},\omega)]^{2}}, ~~~~
\end{eqnarray}
where ${\rm Re}\Sigma_{\rm tot}({\bf k},\omega)$ and
${\rm Im}\Sigma_{\rm tot}({\bf k},\omega)$ are the real and imaginary parts of the total
homogenous self-energy,
\begin{eqnarray}\label{TOT-SE}
\Sigma_{\rm tot}({\bf k},\omega)&=&\Sigma_{0}({\bf k},\omega)+\Sigma_{3}({\bf k},\omega)
\nonumber\\
&+&{\Sigma^{2}_{1}({\bf k},\omega)+\Sigma^{2}_{2}({\bf k},\omega)\over\omega
+\varepsilon_{\bf k}-\Sigma_{0}({\bf k},\omega)+\Sigma_{3}({\bf k},\omega)},
\end{eqnarray}
respectively, and then the homogenous quasiparticle excitation spectrum in the SC-state can
be obtained as,
\begin{eqnarray}\label{BQES}
I({\bf k},\omega)=|M({\bf k},\omega)|^{2}n_{\rm F}(\omega)A({\bf k},\omega),
\end{eqnarray}
with the fermion distribution $n_{\rm F}(\omega)$ and the dipole matrix element
$M({\bf k},\omega)$. However, the important point is that $M({\bf k},\omega)$ does not have
any significant energy or temperature dependence \cite{Damascelli03,Campuzano04,Fink07}. In
this case, the magnitude of $M({\bf k},\omega)$ can be rescaled to the unit, and then the
evolution of $I({\bf k},\omega)$ with momentum, energy, temperature, and doping concentration
is completely characterized by the electron spectral function $A({\bf k},\omega)$.

\begin{figure}[h!]
\centering
\includegraphics[scale=0.045]{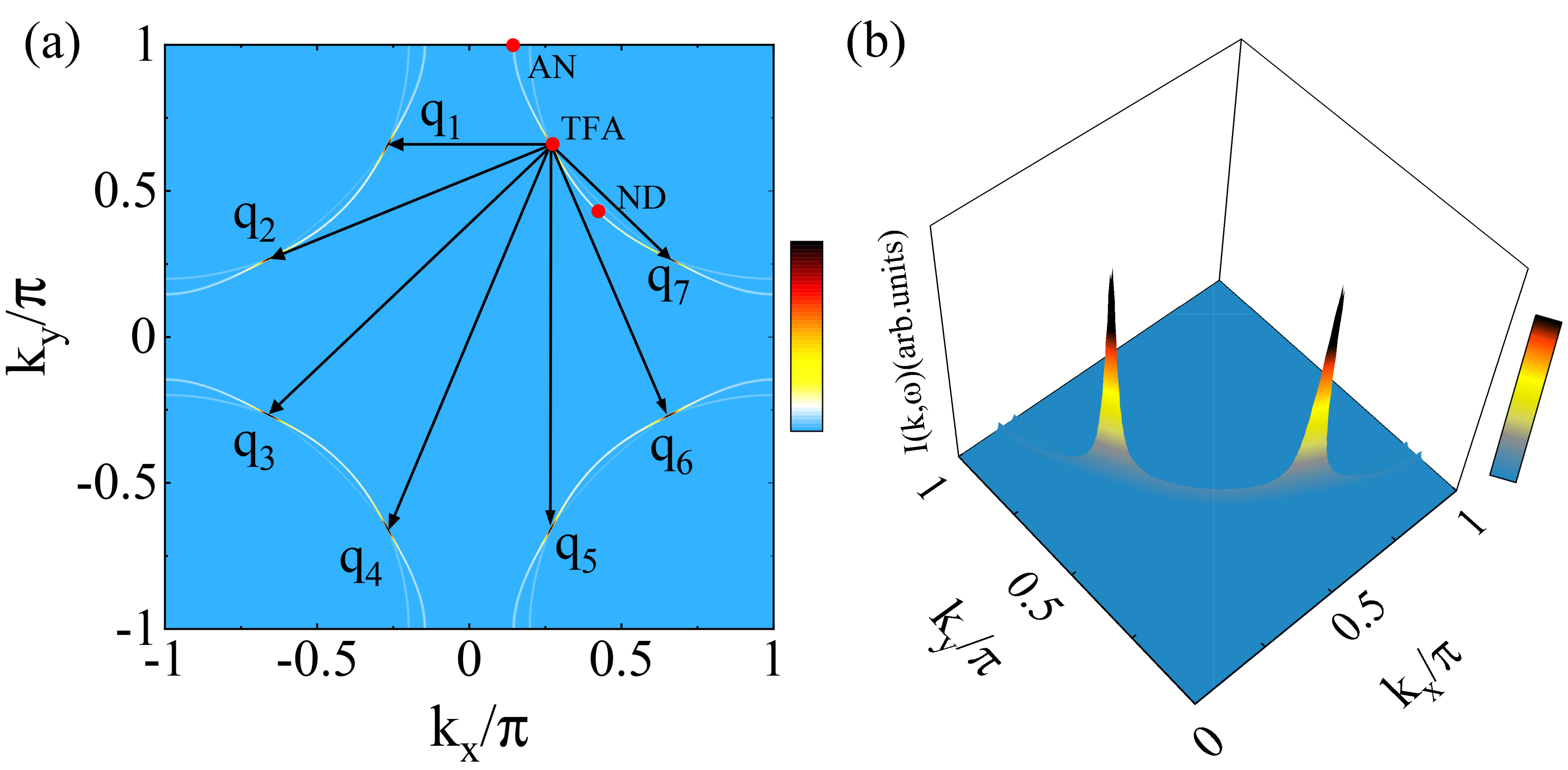}
\caption{(Color online) (a) The electron Fermi surface map and (b) the surface plot of the
homogenous quasiparticle excitation spectrum for zero energy $\omega=0$ at $\delta=0.15$ with
$T=0.002J$, where AN, TFA, and ND denote the antinode, tip of the Fermi arc, and node,
respectively, while ${\bf q}_{1}$, ${\bf q}_{2}$, ${\bf q}_{3}$, ${\bf q}_{4}$,
${\bf q}_{5}$, ${\bf q}_{6}$, and ${\bf q}_{7}$ indicate different scattering wave vectors.
\label{EFS-MAP}}
\end{figure}

In the previous studies \cite{Liu21,Gao18}, the topology of EFS in the pure system has been
discussed in terms of the intensity map of the homogenous quasiparticle excitation spectrum
$I({\bf k},\omega)$ at zero energy $\omega=0$, where we have shown that the formation of the
disconnected Fermi arcs due to the EFS reconstruction is directly associated with the
emergence of the highly anisotropic momentum-dependence of the homogenous quasiparticle
scattering rate. For a convenience in the following discussions of the impurity scattering
influence on the electronic structure, (a) the EFS map in the pure system and (b) the surface
plot of the homogenous quasiparticle excitation spectrum for zero energy $\omega=0$ at doping
$\delta=0.15$ with temperature $T=0.002J$ are {\it replotted} in Fig. \ref{EFS-MAP}.
Obviously, the typical feature is that EFS contour is broken up into the disconnected Fermi
arcs located around the nodal region \cite{Shi08,Sassa11,Comin14,Horio16,Loret18}, where a
large number of the low-energy electronic states is available at around the tips of the Fermi
arcs, and then all the anomalous properties arise from these quasiparticles at around the
tips of the Fermi arcs \cite{Timusk99,Hufner08,Comin16,Vishik18}. These tips of the Fermi
arcs connected by the scattering wave vectors ${\bf q}_{i}$ shown in Fig. \ref{EFS-MAP}
naturally construct an {\it octet scattering model}, and then the quasiparticle scattering
processes with the scattering wave vectors ${\bf q}_{i}$ therefore contribute effectively to
the quasiparticle scattering processes
\cite{Yin21,Pan01,Hoffman02,Kohsaka07,Kohsaka08,Hamidian16}. As we have mentioned in section
\ref{Introduction}, this {\it octet scattering model} shown in Fig. \ref{EFS-MAP} can
persist into the case for a finite binding-energy \cite{Chatterjee06,He14}, which leads to
that the sharp peaks in the ARPES autocorrelation spectrum with the scattering wave vectors
${\bf q}_{i}$ are directly correlated to the regions of the highest joint density of states.
We will return to this discussion of the ARPES autocorrelation towards
Sec. \ref{ARPES-autocorrelation} of this paper.

\subsection{Dressed electron propagator}\label{dressed-propagator}

With the help of the above homogenous electron propagator (\ref{EGF}), now we can discuss
the influence of the impurity scattering on the electronic structure. In the presence of
impurities, the homogenous electron propagator (\ref{EGF}) is dressed via the impurity
scattering as \cite{Hussey02,Balatsky06,Alloul09},
\begin{eqnarray}\label{ID-EGF-1}
\tilde{G}_{\rm I}({\bf k},\omega)^{-1}=\tilde{G}({\bf k},\omega)^{-1}
-\tilde{\Sigma}_{\rm I}({\bf k},\omega),
\end{eqnarray}
where as the homogenous self-energy
$\tilde{\Sigma}({\bf k},\omega)=\sum_{\alpha=0}^{3}\tilde{\Sigma}_{\alpha}({\bf k},\omega)$
in Eq. (\ref{EGF}), the impurity scattering self-energy
$\tilde{\Sigma}_{\rm I}({\bf k},\omega)$ can be also generally expressed as,
\begin{eqnarray}\label{ESE-IS}
&~&\tilde{\Sigma}_{\rm I}({\bf k},\omega)=\sum_{\alpha=0}^{3}
\Sigma_{{\rm I}\alpha}({\bf k},\omega)\tau_{\alpha}\nonumber\\
&=&\left(
\begin{array}{cc}
\Sigma_{\rm I0}({\bf k},\omega)+\Sigma_{\rm I3}({\bf k},\omega),
&\Sigma_{\rm I1}({\bf k},\omega)-i\Sigma_{\rm I2}({\bf k},\omega)\\
\Sigma_{\rm I1}({\bf k},\omega)+i\Sigma_{\rm I2}({\bf k},\omega),
&\Sigma_{\rm I0}({\bf k},\omega)-\Sigma_{\rm I3}({\bf k},\omega)
\end{array}\right).~~~~~
\end{eqnarray}
Moreover, in corresponding to the homogenous self-energies $\Sigma_{1}({\bf k},\omega)$,
$\Sigma_{2}({\bf k},\omega)$, $\Sigma_{3}({\bf k},\omega)$, and $\Sigma_{0}({\bf k},\omega)$
in Eq. (\ref{EGF}), both $\Sigma_{\rm I1}({\bf k},\omega)$ and
$\Sigma_{\rm I2}({\bf k},\omega)$ are real, while both $\Sigma_{\rm I3}({\bf k},\omega)$,
and $\Sigma_{\rm I0}({\bf k},\omega)/\omega$ are an even function of $\omega$. Substituting
this impurity scattering self-energy (\ref{ESE-IS}) and homogenous electron propagator
(\ref{EGF}) into Eq. (\ref{ID-EGF-1}), the dressed electron propagator can be expressed as,
%\begin{widetext}
\begin{eqnarray}\label{ID-EGF}
\tilde{G}_{\rm I}({\bf k},\omega)&=&\left(
\begin{array}{cc}
G_{\rm I}({\bf k},\omega), & \Im_{\rm I}({\bf k},\omega) \\
\Im^{\dagger}_{\rm I}({\bf k},\omega), & -G_{\rm I}({\bf k},-\omega)
\end{array}\right)\nonumber\\
&=&{1\over F_{\rm I}({\bf k},\omega)}\{[\omega-\Sigma_{0}({\bf k},\omega)
-\Sigma_{\rm I0}({\bf k},\omega)]\tau_{0}\nonumber\\
&+&[\Sigma_{1}({\bf k},\omega)+\Sigma_{\rm I1}({\bf k},\omega)]\tau_{1}\nonumber\\
&+&[\Sigma_{2}({\bf k},\omega)+\Sigma_{\rm I2}({\bf k},\omega)]\tau_{2}\nonumber\\
&+&[\varepsilon_{\bf k}+\Sigma_{3}({\bf k},\omega)+\Sigma_{\rm I3}({\bf k},\omega)]
\tau_{3}\},~~
\end{eqnarray}
%\end{widetext}
where $F_{\rm I}({\bf k},\omega)=[\omega-\Sigma_{0}({\bf k},\omega)
-\Sigma_{\rm I0}({\bf k},\omega)]^{2}-[\varepsilon_{\bf k}+\Sigma_{3}({\bf k},\omega)
+\Sigma_{\rm I3}({\bf k},\omega)]^{2}-[\Sigma_{1}({\bf k},\omega)
+\Sigma_{\rm I1}({\bf k},\omega)]^{2}-[\Sigma_{2}({\bf k},\omega)
+\Sigma_{\rm I2}({\bf k},\omega)]^{2}$.

\subsection{Self-consistent T-matrix approach}\label{T-matrix-approach}

\begin{figure}[h!]
\centering
\includegraphics[scale=0.22]{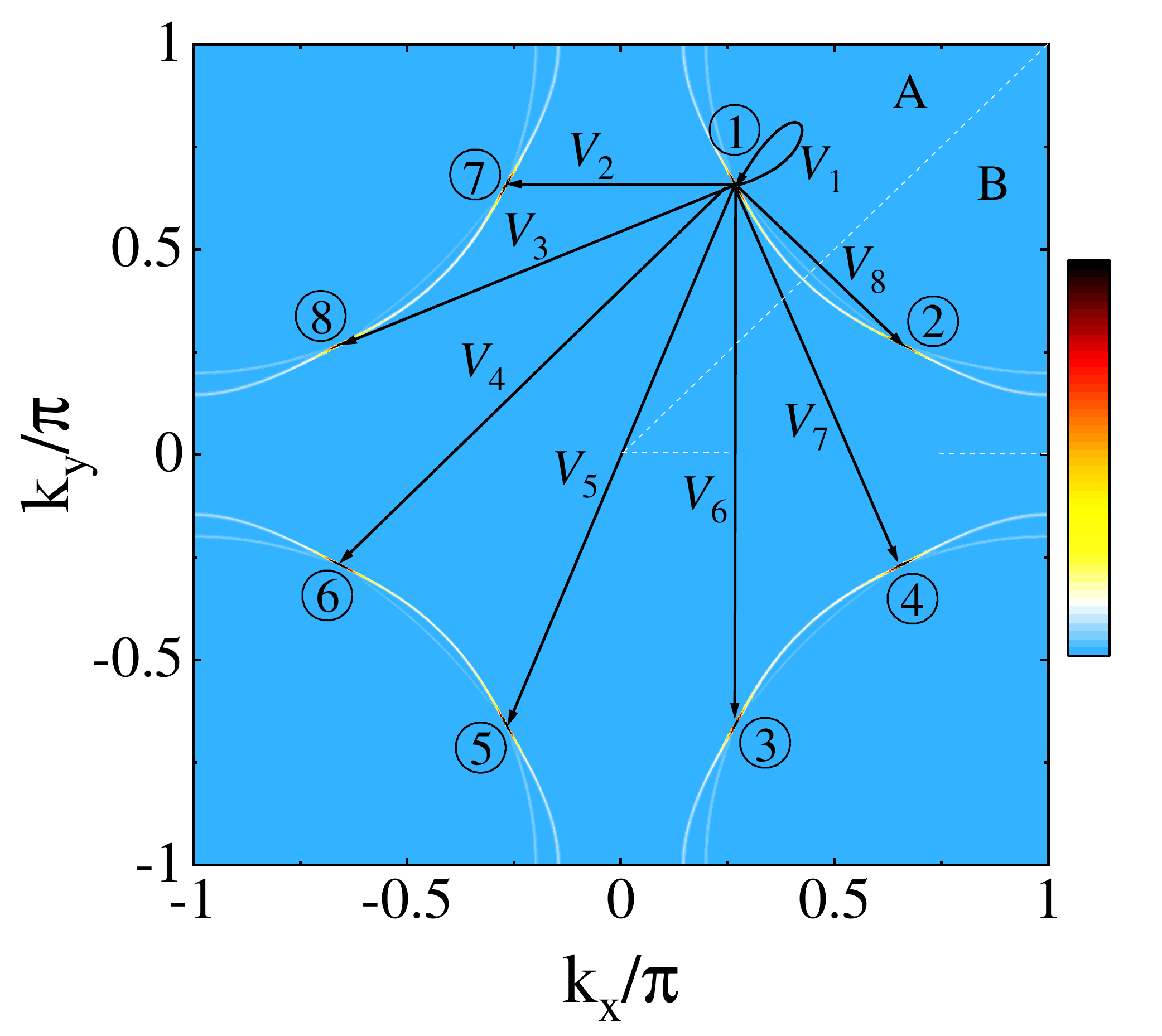}
\caption{(Color online) The impurity scattering in the octet scattering model. $V_{1}$ is the
impurity scattering potential for the intra-tip scattering, $V_{2}$, $V_{3}$, $V_{7}$, and
$V_{8}$ are the impurity scattering potentials for the adjacent-tip scattering, while $V_{4}$,
$V_{5}$, and $V_{6}$ are the impurity scattering potentials for the opposite-tip scattering.
The tips of the Fermi arcs (then the scattering centers) are divided into two groups: (A) the
tips of the Fermi arcs located at the region of $|k_{y}|>|k_{x}|$ and (B) the tips of the Fermi
arcs located at the region of $|k_{x}|>|k_{y}|$. \label{Tip-approximation}}
\end{figure}

Starting from the homogenous part of the BCS-like electron propagator with the d-wave symmetry,
it has been shown that the self-consistent $T$-matrix approach is a powerful tool to treat the
impurity scattering in the SC-state for an arbitrary scattering strength
\cite{Hussey02,Balatsky06,Alloul09,Mahan81,Hirschfeld89,Hirschfeld93}. In the following
discussions, we employ the self-consistent $T$-matrix approach to analyze the impurity
scattering self-energy $\tilde{\Sigma}_{\rm I}({\bf k},\omega)$ in Eq. (\ref{ESE-IS}) in terms
of the dressed electron propagator (\ref{ID-EGF}). Following the self-consistent $T$-matrix
approach \cite{Hussey02,Balatsky06,Alloul09,Mahan81,Hirschfeld89,Hirschfeld93}, the impurity
scattering self-energy (\ref{ESE-IS}) can be expressed approximately as,
\begin{eqnarray}\label{SE-FIS-1}
\tilde{\Sigma}_{\rm I}({\bf k},\omega)=n_{\rm i}N\tilde{T}_{{\bf k}{\bf k}}(\omega),
\end{eqnarray}
where $n_{\rm i}$ is the impurity concentration, $N$ is the number of sites on a square lattice,
and $\tilde{T}_{{\bf k}{\bf k}}(\omega)$ is the diagonal part of the T-matrix, while the T-matrix
is given by the summation of all impurity scattering processes as,
\begin{equation}\label{TMAT_ORI}
\tilde{T}_{{\bf k}{\bf k}'}={1\over N}\tau_{3}V_{{\bf k}{\bf k}'}+{1\over N}
\sum_{{\bf k}''}V_{{\bf k}{\bf k}''}\tau_{3}\tilde{G}_{\rm I}({\bf k}'',\omega)
\tilde{T}_{{\bf k}''{\bf k}'},
\end{equation}
with the impurity scattering potential $V_{{\bf k}{\bf k}'}$, where we have followed the
common practice \cite{Hussey02,Balatsky06,Alloul09,Mahan81,Hirschfeld89,Hirschfeld93}, and
treated the impurity scattering potential $V_{{\bf k}{\bf k}'}$ in the static-limit for a
qualitative understanding of the influence of impurities on the low-energy electronic
structure of cuprate superconductors.

In the octet scattering model shown in Fig. \ref{EFS-MAP}, a large number of the low-energy
electronic states is located at around eight tips of the Fermi arcs. In other words, the
most quasiparticles are generated only at around these tips of the Fermi arcs. This
characteristic feature is very helpful when one considers the impurity scattering, since
the initial and final momenta of a scattering event must always be approximately equal to
the ${\bf k} $-space located at around one of these tips of the Fermi arcs in the case of
low-temperature and low-energy. On the other hand, the impurity scattering potential
$V_{{\bf k}{\bf k}'}$ varies slowly over the area around the tip of the Fermi arc, and
thus the impurity scattering potential can be approximated to be identical within one half
of each quarter in the Brillouin zone (BZ). In this case, a general impurity scattering
potential $V_{{\bf k}{\bf k}'}$ in Eq. (\ref{TMAT_ORI}) need mainly to be considered in
three possible cases as shown in Fig. \ref{Tip-approximation}: (i) the impurity scattering
potential for the scattering at the intra-tip of the Fermi arc $V_{{\bf k}{\bf k}'}=V_{1}$
(${\bf k}$ and ${\bf k}'$ at the same tip of the Fermi arc); (ii) the impurity scattering
potentials for the scattering at the adjacent-tips of the Fermi arcs
$V_{{\bf k}{\bf k}'}=V_{2}$, $V_{{\bf k}{\bf k}'}=V_{3}$, $V_{{\bf k}{\bf k}'}=V_{7}$, and
$V_{{\bf k}{\bf k}'}=V_{8}$ (${\bf k}$ and ${\bf k}'$ at the adjacent-tips of the Fermi
arcs); (iii) and the impurity scattering potentials for the scattering at the opposite-tips
of the Fermi arcs $V_{{\bf k}{\bf k}'}=V_{4}$, $V_{{\bf k}{\bf k}'}=V_{5}$, and
$V_{{\bf k}{\bf k}'}=V_{6}$ (${\bf k}$ and ${\bf k}'$ at the opposite-tips of the Fermi
arcs). This approximation based on the {\it octet scattering model} is so-called as the
{\it Fermi-arc-tip approximation}. It should be emphasized that in this Fermi-arc-tip
approximation, the influence of the impurity scattering on the electron pair strength can
be explored directly, which is much different from the case in the nodal approximation
\cite{Hussey02,Balatsky06,Alloul09}. In this Fermi-arc-tip approximation, the impurity
scattering potential $V_{{\bf k}{\bf k}'}$ in the self-consistent T-matrix equation
(\ref{TMAT_ORI}) is dependent on the momenta at the tips of the Fermi arcs only, and can
be effectively reduced as a $8\times 8$-matrix,
\begin{eqnarray}\label{ISP-matrix}
\tilde{V} =\left(
\begin{array}{cccc}
V_{11} & V_{12} & \cdots & V_{18}\\
V_{21} & V_{22} & \cdots & V_{28}\\
\vdots & \vdots & \ddots & \vdots\\
V_{81} & V_{82} & \cdots & V_{88}
\end{array}\right),
\end{eqnarray}
where the matrix elements are given by: $V_{jj}=V_{1}$ for $j=1,2,3,... 8$,
$V_{jj'}=V_{j'j}=V_{2}$ for $j=1,2,3,6$ with the corresponding $j'=7,4,5,8$, respectively,
$V_{jj'}=V_{j'j}=V_{3}$ for $j=1,2,3,4$ with the corresponding $j'=8,7,6,5$, respectively,
$V_{jj'}=V_{jj'}=V_{4}$ for $j=1,2,3,4$ with the corresponding $j'=6,5,8,7$, respectively,
$V_{jj'}=V_{j'j}=V_{5}$ for $j=1,2,3,4$ with the corresponding $j'=5,6,7,8$, respectively,
$V_{jj'}=V_{j'j}=V_{6}$ for $j=1,2,4,5$ with the corresponding $j'=3,8,6,7$, respectively,
$V_{jj'}=V_{j'j}=V_{7}$ for $j=1,2,5,6$ with the corresponding $j'=4,3,8,7$, respectively,
and $V_{jj'}=V_{j'j}=V_{8}$, for $j=1,3,5,7$ with the corresponding $j'=2,4,6,8$,
respectively.

At the case of zero temperature and zero energy, the Fermi arc collapses to the point at
the tip of the Fermi arc, leading to form the Fermi-arc-tip liquid \cite{Liu21,Gao18}, where
all the spectral weights on the Fermi arc are reduced to the point at the tip of the Fermi
arc, indicating that the quasiparticles are only generated at the tips of the Fermi arcs and
the rest of BZ makes no contribution. In this case, the scattering processes in the octet
scattering model shown in Fig. \ref{Tip-approximation} represent all the scattering
processes in the system, and then in principle, the Fermi-arc-tip approximation for the
impurity scattering potentials can reproduce properly any impurity scattering potential with
arbitrary strength, especially the adjacent scattering potential for the scattering at two
different tips of the Fermi arcs. On the other hand, at the case of low-temperature and
low-energy, although the spectral weight on the point at the tip of the Fermi arc spreads on
the extremely small area around the point at the tip of the Fermi arc, the characteristic
feature of the Fermi arc with the most part of the spectral weight located around the point
at tip of the Fermi arc remains \cite{Chatterjee06,He14,Gao18a,Gao19}, indicating that the
Fermi-arc-tip approximation is still appropriate to treat the impurity scattering at the
case of low-temperature and low-energy. In the following discussions, we therefore employ
the reduced impurity scattering potential (\ref{ISP-matrix}) to study the influence of the
impurity scattering on the electronic structure. Substituting the impurity scattering
potential $\tilde{V}$ in Eq. (\ref{ISP-matrix}) into Eq. (\ref{TMAT_ORI}), the T-matrix
equation can be expressed explicitly as a $16\times 16$-matrix equation around eight tips
of the Fermi arcs as,
\begin{equation}\label{T-matrix-ISP}
\tilde{T}_{jj'}={1\over N}\tau_{3}V_{jj'}+{1\over N}\sum_{j''{\bf k}''}V_{jj''}[\tau_{3}
\tilde{G}_{\rm I}({\bf k}'',\omega)]\tilde{T}_{j''j'},
\end{equation}
where $j$, $j'$, and $j''$ are labels of the tips of the Fermi arcs, the summation
${\bf k}''$ is over the area around the tip $j''$ of the Fermi arc, and then the impurity
scattering self-energy $\tilde{\Sigma}_{\rm I}({\bf k},\omega)$ in Eq. (\ref{SE-FIS-1}) is
reduced as,
\begin{eqnarray}\label{SE-FIS}
\tilde{\Sigma}_{\rm I}(\omega)=n_{\rm i}N\tilde{T}_{jj}(\omega),
\end{eqnarray}
and therefore is also dependent on the momenta at the tips of the Fermi arcs only.

It should be noted that the typical feature of the octet scattering model shown in
Fig. \ref{EFS-MAP} is that two tips of the Fermi arc in each quarter of BZ is symmetrical
about the nodal (diagonal) direction, reflecting a basic fact that the diagonal propagator
in Eq. (\ref{EGF}) is symmetrical about the nodal direction. However, the off-diagonal
propagator in Eq. (\ref{EGF}) is asymmetrical about the nodal direction, since the homogenous
self-energies $\Sigma_{1}({\bf k},\omega)$ and $\Sigma_{2}({\bf k},\omega)$ in the
particle-particle channel (then the momentum and energy dependence of the homogenous SC gap)
have a d-wave symmetry in the framework of the kinetic-energy-driven superconductivity. In
this case, we can divide the region of the location of the tips of the Fermi arcs (then the
scattering centers) into two groups: (A) the tips of the Fermi arcs located at the region of
$|k_{y}|>|k_{x}|$ and (B) the tips of the Fermi arcs located at the region of
$|k_{x}|>|k_{y}|$. Since the symmetry of the impurity scattering self-energy
$\tilde{\Sigma}_{\rm I}(\omega)$ is the same as the homogenous self-energy
$\tilde{\Sigma}({\bf k},\omega)$, the dressed electron propagator
$\tilde{G}_{I}({\bf k},\omega)$ in Eq. (\ref{ID-EGF-1}) can be expressed explicitly in the
regions A and B as,
%\begin{widetext}
\begin{subequations}\label{ID-EGF-AB}
\begin{eqnarray}
\tilde{G}^{\rm (A)}_{\rm I}({\bf k},\omega)&=&\left(
\begin{array}{cc}
G^{\rm (A)}_{\rm I}({\bf k},\omega), & \Im^{\rm (A)}_{\rm I}({\bf k},\omega) \\
\Im^{{\rm (A)}\dagger}_{\rm I}({\bf k},\omega), & -G^{\rm (A)}_{\rm I}({\bf k},-\omega)
\end{array}\right)\nonumber\\
&=&{1\over F^{\rm (A)}_{\rm I}({\bf k},\omega)}\{[\omega-\Sigma_{0}({\bf k},\omega)
-\Sigma_{\rm I0}(\omega)]\tau_{0}\nonumber\\
&+&[\Sigma_{1}({\bf k},\omega)+\Sigma^{\rm (A)}_{\rm I1}(\omega)]\tau_{1}\nonumber\\
&+&[\Sigma_{2}({\bf k},\omega)+\Sigma^{\rm (A)}_{\rm I2}(\omega)]\tau_{2}\nonumber\\
&+&[\varepsilon_{\bf k}+\Sigma_{3}({\bf k},\omega)+\Sigma_{\rm I3}(\omega)]
\tau_{3}\},~~\\
\tilde{G}^{\rm (B)}_{\rm I}({\bf k},\omega)&=&\left(
\begin{array}{cc}
G^{\rm (B)}_{\rm I}({\bf k},\omega), & \Im^{\rm (B)}_{\rm I}({\bf k},\omega) \\
\Im^{{\rm (B)}\dagger}_{\rm I}({\bf k},\omega), & -G^{\rm (B)}_{\rm I}({\bf k},-\omega)
\end{array}\right)\nonumber\\
&=&{1\over F^{\rm (B)}_{\rm I}({\bf k},\omega)}\{[\omega-\Sigma_{0}({\bf k},\omega)
-\Sigma_{\rm I0}(\omega)]\tau_{0}\nonumber\\
&+&[\Sigma_{1}({\bf k},\omega)+\Sigma^{\rm (B)}_{\rm I1}(\omega)]\tau_{1}\nonumber\\
&+&[\Sigma_{2}({\bf k},\omega)+\Sigma^{\rm (B)}_{\rm I2}(\omega)]\tau_{2}\nonumber\\
&+&[\varepsilon_{\bf k}+\Sigma_{3}({\bf k},\omega)+\Sigma_{\rm I3}(\omega)]
\tau_{3}\},~~
\end{eqnarray}
\end{subequations}
%\end{widetext}
respectively, where $F^{\rm (A)}_{\rm I}({\bf k},\omega)=[\omega-\Sigma_{0}({\bf k},\omega)
-\Sigma_{\rm I0}(\omega)]^{2}-[\varepsilon_{\bf k}+\Sigma_{3}({\bf k},\omega)
+\Sigma_{\rm I3}(\omega)]^{2}-[\Sigma_{1}({\bf k},\omega)
+\Sigma^{\rm (A)}_{\rm I1}(\omega)]^{2}-[\Sigma_{2}({\bf k},\omega)
+\Sigma^{\rm (A)}_{\rm I2}(\omega)]^{2}$, $F^{\rm (B)}_{\rm I}({\bf k},\omega)
=[\omega-\Sigma_{0}({\bf k},\omega)
-\Sigma_{\rm I0}(\omega)]^{2}-[\varepsilon_{\bf k}+\Sigma_{3}({\bf k},\omega)
+\Sigma_{\rm I3}(\omega)]^{2}-[\Sigma_{1}({\bf k},\omega)
+\Sigma^{\rm (B)}_{\rm I1}(\omega)]^{2}-[\Sigma_{2}({\bf k},\omega)
+\Sigma^{\rm (B)}_{\rm I2}(\omega)]^{2}$. With the help of the above dressed electron
propagators $\tilde{G}^{\rm (A)}_{I}({\bf k},\omega)$ and
$\tilde{G}^{\rm (B)}_{I}({\bf k},\omega)$, the self-consistent T-matrix equation
(\ref{T-matrix-ISP}) can be further reduced as,
\begin{eqnarray}\label{T-iterr-1}
\tilde{T}_{jj'} &=& {1\over N}V_{jj'}\tau_{3}+{1\over N}\sum_{j''\in {\rm A}} V_{jj''}
[\tau_{3}\tilde{I}^{\rm (A)}_{\tilde{G}}(\omega)]\tilde{T}_{j''j'}\nonumber \\
&+& {1\over N}\sum_{j''\in {\rm B}}V_{jj''}
[\tau_{3}\tilde{I}^{\rm (B)}_{\tilde{G}}(\omega)]\tilde{T}_{j''j'},
\end{eqnarray}
where $\tilde{I}^{\rm (A)}_{\tilde{G}}(\omega)$ and $\tilde{I}^{\rm (B)}_{\tilde{G}}(\omega)$
are the integral propagators, and can be expressed explicitly as,
\begin{eqnarray}
\tilde{I}^{\rm (A)}_{\tilde{G}}(\omega)&=&\sum_{{\bf k}\in {\rm A}}
\tilde{G}^{\rm (A)}_{\rm I}({\bf k},\omega)=\sum\limits_{\alpha=0}^{3}\tau_{\alpha}
I^{\rm (A)}_{\tilde{G}\alpha}(\omega), \\
\tilde{I}^{\rm (B)}_{\tilde{G}}(\omega)&=&\sum_{{\bf k}\in {\rm B}}
\tilde{G}^{\rm (B)}_{\rm I}({\bf k},\omega)=\sum\limits_{\alpha=0}^{3}\tau_{\alpha}
I^{\rm (B)}_{\tilde{G}\alpha}(\omega),~~~
\end{eqnarray}
respectively. To coincide with the separation of the region of the location of the
Fermi-arc tips, the matrix of the impurity scattering potential $\tilde{V}$ in
Eq. (\ref{ISP-matrix}) now can be rearranged in the following way,
\begin{eqnarray}\label{SP-rearrangement}
{1\over N}\tilde{V} =
\left(
\begin{array}{cc}
\bar{V}_{\rm AA} & \bar{V}_{\rm AB}\\
\bar{V}_{\rm BA} & \bar{V}_{\rm BB}
\end{array}\right),
\end{eqnarray}
with the $4\times 4$-matrices of the impurity scattering potentials
$\bar{V}_{\rm AA}$, $\bar{V}_{\rm AB}$, $\bar{V}_{\rm BA}$, and
$\bar{V}_{\rm BB}$ that are given by,
\begin{subequations}
\begin{eqnarray}
\bar{V}_{\rm AA} &=& {1\over N}\left(\begin{array}{cccc}
V_{11} & V_{13} & V_{15} & V_{17}\\
V_{31} & V_{33} & V_{35} & V_{37}\\
V_{51} & V_{53} & V_{55} & V_{57}\\
V_{71} & V_{73} & V_{75} & V_{77}
\end{array}\right),  \\
\bar{V}_{\rm AB} &=&{1\over N} \left(\begin{array}{cccc}
V_{12} & V_{14} & V_{16} & V_{18}\\
V_{32} & V_{34} & V_{36} & V_{38}\\
V_{52} & V_{54} & V_{56} & V_{58}\\
V_{72} & V_{74} & V_{76} & V_{78}
\end{array}\right),  \\
\bar{V}_{\rm BA} &=&{1\over N} \left(\begin{array}{cccc}
V_{21} & V_{23} & V_{25} & V_{27}\\
V_{41} & V_{43} & V_{45} & V_{47}\\
V_{61} & V_{63} & V_{65} & V_{67}\\
V_{81} & V_{83} & V_{85} & V_{87}
\end{array}\right),   \\
\bar{V}_{\rm BB} &=&{1\over N} \left(\begin{array}{cccc}
V_{22} & V_{24} & V_{26} & V_{28}\\
V_{42} & V_{44} & V_{46} & V_{48}\\
V_{62} & V_{64} & V_{66} & V_{68}\\
V_{82} & V_{84} & V_{86} & V_{88}
\end{array}\right), ~~~~~~~~
\end{eqnarray}
\end{subequations}
respectively. According to the above impurity scattering potential $\tilde{V}$ in
Eq. (\ref{SP-rearrangement}), the self-consistent T-matrix equation (\ref{T-iterr-1}) then
can be rewritten as,
\begin{eqnarray}\label{T-matrix-5}
\tilde{T}_{\mu\nu} &=& \bar{V}_{\mu\nu}\otimes\tau_{3}+\bar{V}_{\mu{\rm A}}\otimes
[\tau_{3}\tilde{I}^{\rm (A)}_{\tilde{G}}(\omega)]\tilde{T}_{{\rm A}\nu}\nonumber\\
&+& \bar{V}_{\mu{\rm B}}\otimes [\tau_{3}
\tilde{I}^{\rm (A)}_{\tilde{G}}(\omega)]\tilde{T}_{{\rm B}\nu},
\end{eqnarray}
where $\mu$ ($\nu$) denotes region A or B. After a quite complicated calculation, the above
T-matrix equation now can be evaluated as [see Appendix \ref{T-matrix-equation}],
\begin{eqnarray}\label{T-matrix-6}
\sum\limits_{\alpha=0}^{3}T^{(\alpha)}_{\mu\nu}\otimes\tau_{\alpha}\tau_{3}=
\sum\limits_{\alpha=0}^{3}\big(\sum\limits_{\mu'={\rm A},{\rm B}}\bar{V}_{\mu\mu'}
\bar{\Lambda}^{(\alpha)}_{\mu'\nu}\big)\otimes\tau_{\alpha},
\end{eqnarray}
with the matrix $\bar{\Lambda}^{(\alpha)}$,
\begin{eqnarray}
\bar{\Lambda}^{(\alpha)}\otimes\tau_{\alpha}=\bar{M}={1\over 1-\tilde{M}},
\end{eqnarray}
where the matrix $\tilde{M}$ is obtained as,
\begin{eqnarray}\label{M-matrix}
\tilde{M} &=& \left(
\begin{array}{cc}
\bar{V}_{\rm AA}\otimes\tau_{3}\tilde{I}^{\rm (A)}_{\tilde{G}}(\omega), &
\bar{V}_{\rm AB}\otimes\tau_{3}\tilde{I}^{\rm (A)}_{\tilde{G}}(\omega)\\
\bar{V}_{\rm BA}\otimes\tau_{3}\tilde{I}^{\rm (B)}_{\tilde{G}}(\omega), &
\bar{V}_{\rm BB}\otimes\tau_{3}\tilde{I}^{\rm (B)}_{\tilde{G}}(\omega)\end{array}
\right ),~~~~~
\end{eqnarray}
and then the elements in the matrix $\bar{\Lambda}^{(\alpha)}$ are given by,
\begin{subequations}
\begin{eqnarray}
\Lambda^{(0)}_{\tfrac{i+1}{2} \tfrac{i'+1}{2}} &=&
\frac{1}{2}(\bar{M}_{ii'}+\bar{M}_{i+1 i'+1}),\\
\Lambda^{(3)}_{\tfrac{i+1}{2}\tfrac{i'+1}{2}} &=&
\frac{1}{2}(\bar{M}_{i i'} -\bar{M}_{i+1 i'+1}),\\
\Lambda^{(1)}_{\tfrac{i+1}{2}\tfrac{i'+1}{2}} &=&
\frac{1}{2}(\bar{M}_{i i'+1} +\bar{M}_{i+1 i'}),\\
\Lambda^{(2)}_{\tfrac{i+1}{2} \tfrac{i'+1}{2}} &=&
\frac{i}{2}(\bar{M}_{i i'+1} -\bar{M}_{i+1 i'}),
\end{eqnarray}
\end{subequations}
with $i$ ($i'$)=1,3, 5, $\cdots$, 15. The solution of this
T-matrix equation (\ref{T-matrix-6}) now is given straightforwardly as,
\begin{subequations}\label{T-matrix-10}
\begin{eqnarray}
T_{\mu\nu}^{(0)}(\omega)&=&\sum\limits_{\mu'={\rm A},{\rm B}}\bar{V}_{\mu\mu'}
\bar{\Lambda}^{(3)}_{\mu'\nu}, \\
T_{\mu\nu}^{(1)}(\omega)&=& i\sum\limits_{\mu'={\rm A},{\rm B}}\bar{V}_{\mu\mu'}
\bar{\Lambda}^{(2)}_{\mu'\nu}, \\
T_{\mu\nu}^{(2)}(\omega)&=& -i\sum\limits_{\mu'={\rm A},{\rm B}}\bar{V}_{\mu\mu'}
\bar{\Lambda}^{(1)}_{\mu'\nu}, \\
T_{\mu\nu}^{(3)}(\omega)&=&\sum\limits_{\mu'={\rm A},{\rm B}}\bar{V}_{\mu\mu'}
\bar{\Lambda}^{(0)}_{\mu'\nu}.
\end{eqnarray}
\end{subequations}
Following this solution of the T-matrix equation, the impurity scattering self-energy
$\tilde{\Sigma}_{\rm I}(\omega)=\sum\limits_{\alpha=0}^{3}\Sigma_{{\rm I}\alpha}(\omega)
\tau_{\alpha}=n_{i}N\tilde{T}_{jj}(\omega)$ in the region A can be obtained as,
\begin{subequations}\label{Self-Energy-A}
\begin{eqnarray}
\Sigma^{\rm (A)}_{\rm I0}(\omega)&=&n_{i}N(T_{\rm AA}^{(0)})_{11}
=Nn_{i}(\sum\limits_{\mu'={\rm A},{\rm B}}\bar{V}_{{\rm A}\mu'}
\bar{\Lambda}^{(3)}_{\mu'{\rm A}})_{11}, ~~~~~~\\
\Sigma^{\rm (A)}_{\rm I1}(\omega)&=& n_{i}N{\rm Re}(T_{\rm AA}^{(1)})_{11}\nonumber\\
&=&-Nn_{i}{\rm Im}(\sum\limits_{\mu'={\rm A},{\rm B}}\bar{V}_{{\rm A}\mu'}
\bar{\Lambda}^{(2)}_{\mu'{\rm A}})_{11}, \\
\Sigma^{\rm (A)}_{\rm I2}(\omega)&=& n_{i}N{\rm Re}(T_{\rm AA}^{(2)})_{11}\nonumber\\
&=&Nn_{i}{\rm Im}(\sum\limits_{\mu'={\rm A},{\rm B}}\bar{V}_{{\rm A}\mu'}
\bar{\Lambda}^{(1)}_{\mu'{\rm A}})_{11}, ~~~~~\\
\Sigma^{\rm (A)}_{\rm I3}(\omega)&=& n_{i}N(T_{\rm AA}^{(3)})_{11}
=Nn_{i}(\sum\limits_{\mu'={\rm A},{\rm B}}\bar{V}_{{\rm A}\mu'}
\bar{\Lambda}^{(0)}_{\mu'{\rm A}})_{11},~~~~~
\end{eqnarray}
\end{subequations}
and in the region B is given by,
\begin{subequations}\label{Self-Energy-B}
\begin{eqnarray}
\Sigma^{\rm (B)}_{\rm I0}(\omega)&=&n_{i}N(T_{\rm BB}^{(0)})_{11}
=Nn_{i}(\sum\limits_{\mu'={\rm A},{\rm B}}\bar{V}_{{\rm B}\mu'}
\bar{\Lambda}^{(3)}_{\mu'{\rm B}})_{11}, ~~~~~~~\\
\Sigma^{\rm (B)}_{\rm I1}(\omega)&=& n_{i}N{\rm Re}(T_{\rm BB}^{(1)})_{11}\nonumber\\
&=&-Nn_{i}{\rm Im}(\sum\limits_{\mu'={\rm A},{\rm B}}\bar{V}_{{\rm B}\mu'}
\bar{\Lambda}^{(2)}_{\mu'{\rm B}})_{11},
\end{eqnarray}
\begin{eqnarray}
\Sigma^{\rm (B)}_{\rm I2}(\omega)&=& n_{i}N{\rm Re}(T_{\rm BB}^{(2)})_{11}\nonumber\\
&=&Nn_{i}{\rm Im}(\sum\limits_{\mu'={\rm A},{\rm B}}\bar{V}_{{\rm B}\mu'}
\bar{\Lambda}^{(1)}_{mu'{\rm B}})_{11}, ~~~~~\\
\Sigma^{\rm (B)}_{\rm I3}(\omega)&=& n_{i}N(T_{\rm BB}^{(3)})_{11}
=Nn_{i}(\sum\limits_{\mu'={\rm A},{\rm B}}\bar{V}_{{\rm B}\mu'}
\bar{\Lambda}^{(0)}_{\mu'{\rm B}})_{11}.~~~~~~~~
\end{eqnarray}
\end{subequations}
The above impurity scattering self-energies in Eqs. (\ref{Self-Energy-A})
and (\ref{Self-Energy-B}) are obtained firstly in the {\it Fermi-arc-tip approximation} of
the quasiparticle excitations and scattering processes based on a microscopic {\it octet}
scattering model.

Since the self-energy in the particle-hole channel is symmetrical about the
nodal direction and the self-energy in the particle-particle channel is asymmetrical about
the nodal direction, the above impurity scattering self-energies in the regions A and B
can be rewritten uniformly as,
\begin{subequations}
\begin{eqnarray}\label{Self-Energy}
\Sigma^{\rm (A)}_{\rm I0}(\omega)&=&\Sigma^{\rm (B)}_{\rm I0}(\omega)
=\Sigma_{\rm I0}(\omega),\\
\Sigma^{\rm (A)}_{\rm I1}(\omega)&=&-\Sigma^{\rm (B)}_{\rm I1}(\omega)
=\Sigma_{\rm I1}(\omega),\\
\Sigma^{\rm (A)}_{\rm I2}(\omega)&=&-\Sigma^{\rm (B)}_{\rm I2}(\omega)
=\Sigma_{\rm I2}(\omega),\\
\Sigma^{\rm (A)}_{\rm I3}(\omega)&=&\Sigma^{\rm (B)}_{\rm I3}(\omega)
=\Sigma_{\rm I3}(\omega),
\end{eqnarray}
\end{subequations}
and then the dressed quasiparticle excitation spectrum now can be obtained as,
\begin{eqnarray}\label{IQES}
I_{\rm I}({\bf k},\omega)=|M({\bf k},\omega)|^{2}n_{\rm F}(\omega)A_{\rm I}({\bf k},\omega),
\end{eqnarray}
where the dressed electron spectral function $A_{\rm I}({\bf k},\omega)$ is obtained
directly from the dressed electron propagator (\ref{ID-EGF}) as,
\begin{eqnarray}\label{IDESF}
&&A_{\rm I}({\bf k},\omega)=-2{\rm Im}G_{\rm I}({\bf k},\omega)\nonumber\\
&=&{-2{\rm Im}\Sigma^{\rm (IM)}_{\rm tot}({\bf k},\omega)\over [\omega-\varepsilon_{\bf k}
-{\rm Re}\Sigma^{\rm (IM)}_{\rm tot}({\bf k},\omega)]^{2}
+[{\rm Im}\Sigma^{\rm (IM)}_{\rm tot}({\bf k},\omega)]^{2}},~~~~~~~~~
\end{eqnarray}
with ${\rm Re}\Sigma^{\rm (IM)}_{\rm tot}({\bf k},\omega)$ and
${\rm Im}\Sigma^{\rm (IM)}_{\rm tot}({\bf k},\omega)$ that are the real and imaginary parts
of the total dressed self-energy,
\begin{eqnarray}\label{ID-TOT-SE}
\Sigma^{\rm (IM)}_{\rm tot}({\bf k},\omega)&=&\Sigma_{\rm ph}({\bf k},\omega)
+\Sigma^{\rm (I)}_{\rm ph}(\omega)\nonumber\\
&+&{|\Sigma_{\rm pp}({\bf k},\omega)+(-1)^{\mu+1}\Sigma^{\rm (I)}_{\rm pp}(\omega)|^{2}\over
\omega+\varepsilon_{\bf k}+\Sigma_{\rm ph}({\bf k},-\omega)
+\Sigma^{\rm (I)}_{\rm ph}(-\omega)},~~~~~~~
\end{eqnarray}
respectively, where $\mu=1,2$ for the regions A and B, respectively,
$\Sigma^{\rm (I)}_{\rm ph}(\omega)=\Sigma_{\rm I0}(\omega)+\Sigma_{\rm I3}(\omega)$ and
$\Sigma^{\rm (I)}_{\rm pp}(\omega)=\Sigma_{\rm I1}(\omega)-i\Sigma_{\rm I2}(\omega)$ are the
impurity scattering self-energies in the particle-hole and particle-particle channels,
respectively. In the previous studies based on the nodal approximation of the quasiparticle
excitations and scattering processes \cite{Hussey02,Balatsky06,Alloul09}, the reasonable
strengths of the intra-node impurity scattering, the adjacent-node impurity scattering, and
the opposite-node impurity scattering have been used to discuss the influence of the impurity
scattering on various properties of the SC-state in cuprate superconductors
\cite{Durst00,Yashenkin01,Nunner05,Dahm05,Wang08,Andersen08,Wang09,Hone17}. Unless otherwise
indicated, the strengths of the intra-tip impurity scattering $V_{1}$, the adjacent-tip
impurity scattering $V_{2}$, $V_{3}$, $V_{7}$, and $V_{8}$, and the opposite-tip impurity
scattering $V_{4}$, $V_{5}$, and $V_{6}$ in the following discussions are chosen as
$V_{1}=58J$, $V_{2}=0.85V_{1}$, $V_{3}=0.8V_{1}$, $V_{7}=0.85V_{1}$, $V_{8}=0.9V_{1}$,
$V_{4}=0.7V_{1}$, $V_{5}=0.65V_{1}$, and $V_{6}=0.75V_{1}$, respectively, to compare with the
previous discussions in the nodal approximation of the quasiparticle excitations and scattering
processes \cite{Durst00,Wang08}.

\section{Quantitative characteristics}\label{Quantitative-characteristics}

The studies of the influence of the impurity scattering on the electronic structure can
offer insight into the fundamental aspects of the quasiparticle excitation in cuprate
superconductors \cite{Hussey02,Balatsky06,Alloul09}, and therefore also can offer points
of the reference against which theories may be compared. In this section, we analyze the
quantitative characteristics of the influence of the impurity scattering on the electronic
structure of cuprate superconductors in the SC-state to shed light on the nature of the
SC-state quasiparticle excitation.

\subsection{Impurity concentration dependence of impurity scattering self-energy}
\label{ESEFIS}

\begin{figure}[h!]
\centering
\includegraphics[scale=0.04]{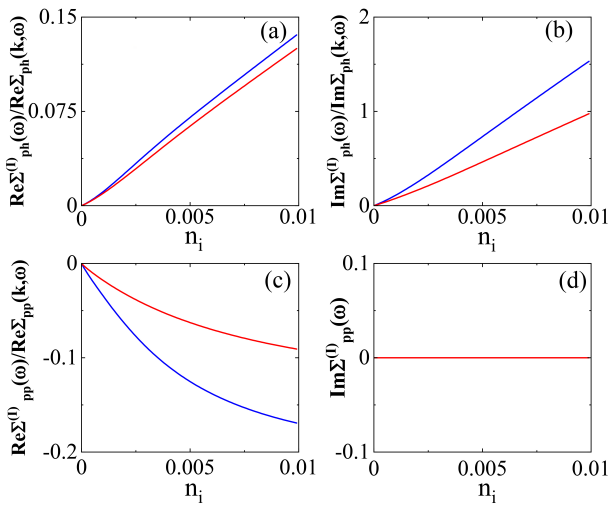}
\caption{(Color online) (a) The real part and (b) the imaginary part of the impurity
scattering self-energy in the particle-hole channel and (c) the real part and (d) the
imaginary part of the impurity scattering self-energy in the particle-particle channel at the
antinode as a function of the impurity concentration at $\delta=0.15$ with $T=0.002J$ in zero
energy $\omega=0$ for the strengths of the adjacent-tip impurity scattering $V_{2}=0.85V_{1}$,
$V_{3}=0.8V_{1}$, $V_{7}=0.85V_{1}$, and $V_{8}=0.9V_{1}$, and the opposite-tip impurity
scattering $V_{4}=0.7V_{1}$, $V_{5}=0.65V_{1}$, and $V_{6}=0.75V_{1}$, and the intra-tip
impurity scattering $V_{1}=58J$ (blue-line) and $V_{1}=30J$ (red-line).
${\rm Re}\Sigma_{\rm ph}({\bf k},\omega)$ and ${\rm Im}\Sigma_{\rm ph}({\bf k},\omega)$ are
the corresponding real and imaginary parts of the homogenous self-energy in the particle-hole
channel, while ${\rm Re}\Sigma_{\rm pp}({\bf k},\omega)$ is the corresponding real part of
the homogenous self-energy in the particle-particle channel. \label{Real-Imaginary-Part}}
\end{figure}

In the framework of the kinetic-energy-driven superconductivity
\cite{Feng0306,Feng12,Feng15a,Feng15}, the electrons interact strongly with spin excitations
resulting in the formation of the quasiparticles, and then all the unconventional features
in the pure cuprate superconductors are mainly dominated by these quasiparticle behaviors
\cite{Timusk99,Hufner08,Comin16,Vishik18}. The quasiparticle energy and lifetime in the pure
system are mainly determined by the real and imaginary parts of the homogenous self-energy
in the particle-hole channel, respectively, while the homogenous self-energy in the
particle-particle channel is identified as the energy and momentum dependence of the
homogenous SC gap in the quasiparticle excitation spectrum, and therefore is corresponding
to the energy for breaking an electron pair. However, the coupling between these
quasiparticles in the pure system and impurities leads to a further renormalization of both
the energy and lifetime of the quasiparticles. To see this further renormalization more
clearly, we firstly analyze the characteristic features of the impurity concentration
dependence of the impurity scattering self-energy. In Fig. \ref{Real-Imaginary-Part}, we
plot (a) the real part ${\rm Re}\Sigma^{\rm (I)}_{\rm ph}(\omega)$ and (b) the imaginary
part ${\rm Im}\Sigma^{\rm (I)}_{\rm ph}(\omega)$ of the impurity scattering self-energy
$\Sigma^{\rm (I)}_{\rm ph}(\omega)$ in the particle-hole channel and (c) the real part
${\rm Re}\Sigma^{\rm (I)}_{\rm pp}(\omega)$ and (d) the imaginary part
${\rm Im}\Sigma^{\rm (I)}_{\rm pp}(\omega)$ of the impurity scattering self-energy
$\Sigma^{\rm (I)}_{\rm pp}(\omega)$ in the particle-particle channel at the antinode as a
function of the impurity concentration at $\delta=0.15$ with $T=0.002J$ in zero energy
$\omega=0$ for the strengths of the adjacent-tip impurity scattering $V_{2}=0.85V_{1}$,
$V_{3}=0.8V_{1}$, $V_{7}=0.85V_{1}$, and $V_{8}=0.9V_{1}$, and the opposite-tip impurity
scattering $V_{4}=0.7V_{1}$, $V_{5}=0.65V_{1}$, and $V_{6}=0.75V_{1}$, and the intra-tip
impurity scattering $V_{1}=58J$ (blue-line) and $V_{1}=30J$ (red-line). The main features of
the impurity scattering self-energy in Fig. \ref{Real-Imaginary-Part} can be summarized as:
(i) the values of both
${\rm Re}\Sigma^{\rm (I)}_{\rm ph}(\omega)/{\rm Re}\Sigma_{\rm ph}({\bf k},\omega)$ and
${\rm Im}\Sigma^{\rm (I)}_{\rm ph}(\omega)/{\rm Im}\Sigma_{\rm ph}({\bf k},\omega)$ are
positive, indicating that the binding-energy in the pure system is shifted by
${\rm Re}\Sigma^{\rm (I)}_{\rm ph}(\omega)$ and the dispersion is further broadened by
${\rm Im}\Sigma^{\rm (I)}_{\rm ph}(\omega)$, where ${\rm Re}\Sigma_{\rm ph}({\bf k},\omega)$
and ${\rm Im}\Sigma_{\rm ph}({\bf k},\omega)$ are the corresponding real and imaginary parts
of the homogenous self-energy in the particle-hole channel.
In particular, with the increase of the impurity concentration, the magnitudes of both
${\rm Re}\Sigma^{\rm (I)}_{\rm ph}(\omega)/{\rm Re}\Sigma_{\rm ph}({\bf k},\omega)$ and
${\rm Im}\Sigma^{\rm (I)}_{\rm ph}(\omega)/{\rm Im}\Sigma_{\rm ph}({\bf k},\omega)$ are
linearly raised \cite{Vobornik00}, which leads to a linear suppression of the spectral
weight of the quasiparticle excitation spectrum and a linear reduction of the lifetime of
the quasiparticle \cite{Vobornik99,Vobornik00,Shen04,Kondo07,Pan09}. (ii) for an any given
impurity concentration, although the magnitude of
${\rm Im}\Sigma^{\rm (I)}_{\rm pp}(\omega)$ is equal to zero,
${\rm Re}\Sigma^{\rm (I)}_{\rm pp}(\omega)/{\rm Re}\Sigma_{\rm pp}({\bf k},\omega)$ has a
negative value, where ${\rm Re}\Sigma_{\rm pp}({\bf k},\omega)$ is the corresponding real
part of the homogenous self-energy in the particle-particle channel. However, the absolute
value of ${\rm Re}\Sigma^{\rm (I)}_{\rm pp}(\omega)/{\rm Re}\Sigma_{\rm pp}({\bf k},\omega)$
is found to monotonically increase as the impurity concentration is increased. The
kinetic-energy-driven SC-state in the pure system \cite{Feng15a} is characterized by the
d-wave SC gap $\bar{\Delta}_{\rm d}({\bf k},\omega)=\Sigma_{\rm pp}({\bf k},\omega)
=\bar{\Delta}_{\rm d}(\omega)[{\rm cos}k_{x}-{\rm cos}k_{y}]/2$, which crosses through zero
at each of four nodes on EFS ($k_{x}=\pm k_{y}$). However, the present result of
${\rm Re}\Sigma^{\rm (I)}_{\rm pp}(\omega)/{\rm Re}\Sigma_{\rm pp}({\bf k},\omega)$ in
Fig. \ref{Real-Imaginary-Part}c therefore also indicates that in addition to the d-wave
component of the SC gap $\bar{\Delta}_{\rm d}({\bf k},\omega)$, the isotropic s-wave
component of the gap \cite{Lee93,Franz96}
$\bar{\Delta}^{\rm (I)}_{\rm s}(\omega)=\Sigma^{\rm (I)}_{\rm pp}(\omega)$ is generated by
the impurity scattering potential (\ref{ISP-matrix}), in which the impurities modulate
the pair interaction locally. This mixed gap
$\bar{\Delta}_{\rm mix}({\bf k},\omega)=\bar{\Delta}_{\rm d}({\bf k},\omega)
+(-1)^{\mu+1}\bar{\Delta}^{\rm (I)}_{\rm s}(\omega)$ therefore leads to a coexistence of
the d-wave component of the gap $\bar{\Delta}_{\rm d}({\bf k},\omega)$ and the isotropic
s-wave component of the gap $\bar{\Delta}^{\rm (I)}_{\rm s}(\omega)$ in the SC-state,
where $\mu=1,2$ for the regions A and B of BZ shown in Fig. \ref{Tip-approximation},
respectively. In particular, the behaviour of this mixed gap naturally deviates from the
d-wave behaviour of the SC gap \cite{Vobornik99,Vobornik00,Shen04,Kondo07,Pan09}. In this
case, the increase of the absolute value of
${\rm Re}\Sigma^{\rm (I)}_{\rm pp}(\omega)/{\rm Re}\Sigma_{\rm pp}({\bf k},\omega)$ at
around the antinodal region upon more impurities is nothing, but the smoothly decrease of
the mixed gap $\bar{\Delta}_{\rm mix}({\bf k},\omega)$ in the magnitude at around the
antinodal region, in agreement with the experimental observations \cite{Pan09}. More
importantly, we have also found that the isotropic s-wave component of the gap at around
the nodal region presents a similar impurity concentration dependent behavior at around the
antinodal region shown in Fig. \ref{Real-Imaginary-Part}c, which leads to the opening of
the gap at around the nodal region, with the magnitude that gradually increases with the
increase of the impurity concentration, indicating the existence of a finite gap over the
entire EFS, and also in agreement with the experimental observations \cite{Shen04}. (iii)
apart from the results shown in Fig. \ref{Real-Imaginary-Part}, we have also made a series
of calculations for the impurity scattering self-energy with other different sets of the
strength of the impurity scattering, and these results together with the results shown in
Fig. \ref{Real-Imaginary-Part} therefore indicate that at an any given impurity
concentration, the magnitudes of ${\rm Re}\Sigma^{\rm (I)}_{\rm ph}(\omega)$ and
${\rm Im}\Sigma^{\rm (I)}_{\rm ph}(\omega)$ and the absolute value of
${\rm Re}\Sigma^{\rm (I)}_{\rm pp}(\omega)$ increase with the increase of the strength of
the impurity scattering, which leads to that except for the increase in the strength of the
impurity-induced renormalization of both the energy and lifetime of the quasiparticles, the
extent of the admixing of the d-wave and the isotropic s-wave components of the gap is also
strongly extended. These strong impurity concentration dependence of the impurity scattering
self-energy in the particle-hole channel and the coexistence of the d-wave and the isotropic
s-wave components of the gap in the particle-particle channel therefore significantly affect
the nature of the quasiparticle excitation in the pure cuprate superconductors
\cite{Hussey02,Balatsky06,Alloul09}.

\subsection{Impurity concentration dependence of line-shape}

\begin{figure}[h!]
\centering
\includegraphics[scale=0.6]{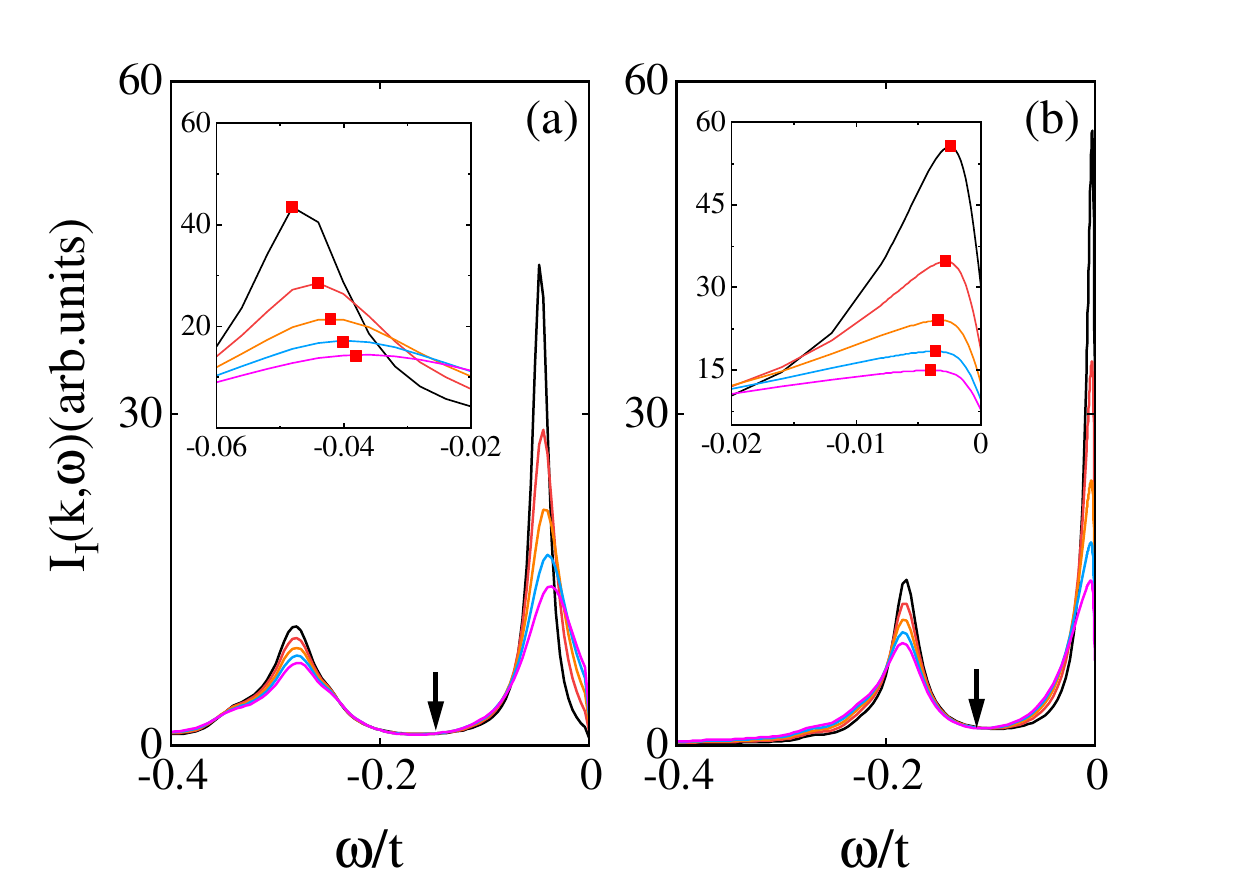}
\caption{(Color online) The dressed quasiparticle excitation spectrum as a function of
energy at (a) the antinode and (b) the node in $\delta=0.15$ with $T=0.002J$ for the
impurity concentrations $n_{i}=0$ (black line), $0.0025$ (red line), $0.005$ (orange
line), $0.0075$ (blue line), and $0.01$ (magenta line), where the arrows indicate the
positions of the dip. The insets in (a) and (b) display the corresponding evolution of
the low binding-energy coherent peaks with the impurity concentration in more detail.
\label{PDH-Spetrum}}
\end{figure}

To reveal how the impurity scattering affects the ARPES spectrum is important to understand
how the quasiparticle excitation behaviour is significantly affected by the impurity
scattering \cite{Hussey02,Balatsky06,Alloul09}. One of the most characteristic features in
the ARPES spectrum of cuprate superconductors is the so-called peak-dip-hump (PDH) structure
\cite{Dessau91,Hwu91,Randeria95,Fedorov99,Lu01,Sakai13,Loret17,DMou17}, which consists of a
coherent peak at the low binding-energy, a broad hump at the higher binding-energy, and a
spectral dip between them. This striking PDH structure has been identified along the entire
EFS \cite{Dessau91,Hwu91,Randeria95,Fedorov99,Lu01,Sakai13,Loret17,DMou17}, and now is a
hallmark of the spectral line-shape of the ARPES spectrum
\cite{Damascelli03,Campuzano04,Fink07}. In particular, the recent ARPES experimental
observations also demonstrate that the same interaction of the electrons with a bosonic
excitation that induces the SC-state in the particle-particle channel also generate a
notable peak structure in the imaginary part of the self-energy in the particle-hole channel
\cite{DMou17}, and then this peak structure induces the remarkable PDH structure in the
ARPES spectrum. Moreover, we \cite{Gao18a} have shown within the framework of the
kinetic-energy-driven superconductivity that this strong coupling of the electrons with the
bosonic excitation can be identified as the strong electron's coupling to a strongly
dispersive spin excitation.
However, the impurity scattering has an important influence on the homogenous self-energies
in the particle-hole and particle-particle channels as we have mentioned in the above
subsection \ref{ESEFIS}, which therefore naturally induces the significant influence on the
intrinsic features of the ARPES spectrum in the pure cuprate superconductors. To see this
significant influence more clearly, we plot the dressed quasiparticle excitation spectrum
$I_{\rm I}({\bf k},\omega)$ as a function of energy at (a) the antinode and (b) the node in
$\delta=0.15$ with $T=0.002J$ for the impurity concentrations $n_{i}=0$ (black-line),
$0.0025$ (red-line), $0.005$ (orange-line), $0.0075$ (blue-line), and $0.01$ (magenta-line)
in Fig. \ref{PDH-Spetrum}, where the spectral signature of the dressed quasiparticle
excitation spectrum is a coherent peak at the low binding-energy, followed by a dip and a
broad hump at the higher binding-energies, in agreement with the ARPES experimental
results \cite{Vobornik99,Vobornik00}.
In the ARPES experiments
\cite{Damascelli03,Campuzano04,Fink07}, a quasiparticle with a long lifetime is observed
as a sharp peak in intensity, and a quasiparticle with a short lifetime is observed as a
broad hump. The results in Fig. \ref{PDH-Spetrum} therefore show clearly that the impurity
scattering induces a broadening of the spectral line together with a shift of the position
of the peak \cite{Vobornik99,Vobornik00,Shen04,Kondo07,Pan09}, i.e., (i) both the coherent
peak at the low binding-energy and the broad hump at the higher binding-energy are
progressively broadened as the impurity concentration increases \cite{Vobornik99,Vobornik00},
leading to the dramatic loss of the intensity of the low binding-energy coherent peak
\cite{Vobornik99,Vobornik00,Shen04,Kondo07,Pan09}. In particular, the progressively loss
of the intensity of the low binding-energy coherent peak with the increase of the impurity
concentration may induce a reduction of $T_{\rm c}$ as that observed in the experiments
\cite{Vobornik99}; (ii) as a natural result of the evolution of the impurity-induced
isotropic s-wave gap $\bar{\Delta}^{\rm (I)}_{\rm s}(\omega)$ with the impurity
concentration obtained in the above Sec. \ref{ESEFIS}, although the position of the dip at
different impurity concentrations is almost invariable, the position of the low
binding-energy coherent peak at around the antinodal region is shifted smoothly towards to
EFS when the impurity concentration is increased \cite{Vobornik00}, while the position of
the low binding-energy coherent peak at around the nodal region progressively moves away
from EFS \cite{Shen04}, also in agreement with the corresponding experimental results
\cite{Vobornik00,Shen04}.

The emergence of the PDH structure in the quasiparticle excitation spectrum can be
attributed to the notable peak structure in the quasiparticle scattering rate originated
from the interaction between electrons by the exchange of spin excitations except for the
impurity-induced a broadening of the spectral line together with a shift of the position of
the coherent peak at the low binding-energy. As the case in the pure system \cite{Gao18a},
the position of the quasiparticle peak in the dressed quasiparticle excitation spectrum
$I_{\rm I}({\bf k},\omega)$ in Eq. (\ref{IQES}) is mainly dominated by the real part of the
total dressed self-energy ${\rm Re}\Sigma^{\rm (IM)}_{\rm tot}({\bf k},\omega)$ in terms of
the following equation,
\begin{eqnarray*}
\omega-\varepsilon_{\bf k}-{\rm Re}\Sigma^{\rm (IM)}_{\rm tot}({\bf k},\omega)=0,
\end{eqnarray*}
and then the lifetime of the quasiparticle at the energy $\omega$ is completely determined
by the inverse of the dressed quasiparticle scattering rate $\Gamma_{\rm I}({\bf k},\omega)$,
which is defined as the imaginary part of the total dressed self-energy as
$\Gamma_{\rm I}({\bf k},\omega)=|{\rm Im}\Sigma^{\rm (IM)}_{\rm tot}({\bf k},\omega)|$.
\begin{figure}[h!]
\centering
\includegraphics[scale=0.57]{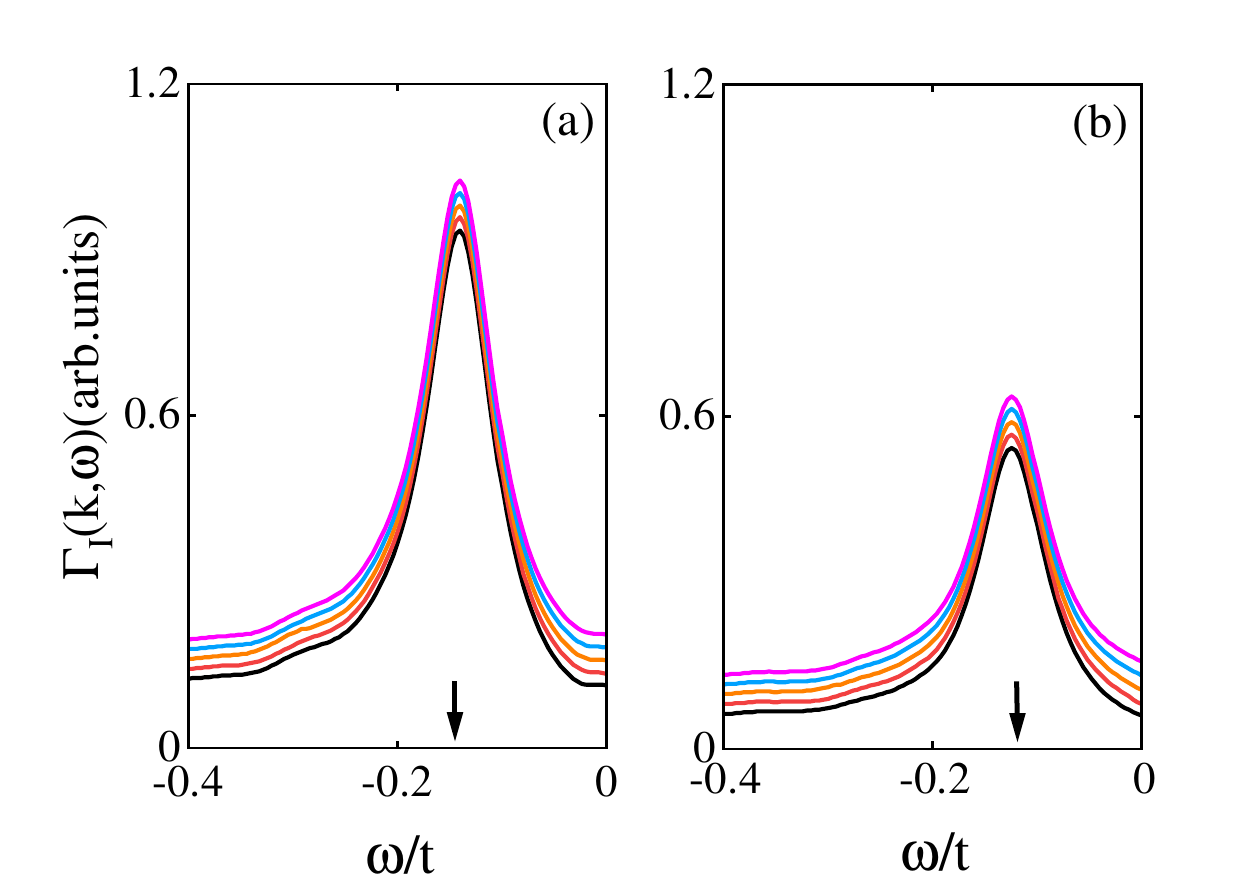}
\caption{(Color online) The dressed quasiparticle scattering rate at (a) the antinode and
(b) the node as a function of energy in $\delta=0.15$ with $T=0.002J$ for the impurity
concentrations $n_{i}=0$ (black-line), $0.0025$ (red-line), $0.005$ (orange-line), $0.0075$
(blue-line), and $0.01$ (magenta-line), where the red arrows indicate the positions of the
peaks. \label{Scattering-Rate}}
\end{figure}
To see this picture more clearly, we plot $\Gamma_{\rm I}({\bf k},\omega)$ as a function of
energy at (a) the antinode and (b) the node in $\delta=0.15$ with $T=0.002J$ for the
impurity concentrations $n_{i}=0$ (black line), $0.0025$ (red line), $0.005$ (orange
line), $0.0075$ (blue line), and $0.01$ (magenta line) in Fig. \ref{Scattering-Rate}. It
thus shows clearly that as the case in the pure system \cite{Gao18a}, the peak structure
also appears at around the antinodal and nodal regions in the presence of the impurity
scattering, where $\Gamma_{\rm I}({\bf k},\omega)$ achieves a sharp peak at the peak
energy, and then it decreases rapidly away from this peak energy \cite{DMou17}. More
importantly, the position of this sharp peak is just corresponding to the position of the
dip in the PDH structure in the dressed quasiparticle excitation spectrum shown in
Fig. \ref{PDH-Spetrum}. In this case, the spectral weight at around the dip energy is
suppressed heavily by the strong quasiparticle scattering, and then the PDH structure is
developed at around the antinodal and nodal regions \cite{DMou17}. On the other hand, the
impurity scattering self-energy in the particle-hole channel further enhances the
quasiparticle scattering as shown in Fig. \ref{Scattering-Rate}, which therefore leads to
a further depression of the spectral weights of the coherent peak at the low binding-energy
and the hump at the higher binding-energy \cite{Vobornik99,Vobornik00,Shen04,Kondo07,Pan09}.
However, the impurity scattering self-energy in the particle-particle channel induces a
strong deviation from the d-wave behaviour of the SC gap (then an existence of a finite gap
over the entire EFS) \cite{Shen04,Kondo07,Pan09} with the exotic impurity concentration
dependence of the gap behaviours at around the nodal and antinodal regions
\cite{Vobornik99,Vobornik00} as we have mentioned in subsection \ref{ESEFIS}, which
thus leads to that with the increase of the impurity concentration, the position of the
low binding-energy coherent peak at around the antinodal region is shifted smoothly towards
to EFS, while the position of the low binding-energy coherent peak at around the nodal
region progressively moves away from EFS.

\subsection{ARPES autocorrelation}\label{ARPES-autocorrelation}

\begin{figure}[h!]
\centering
\includegraphics[scale=0.15]{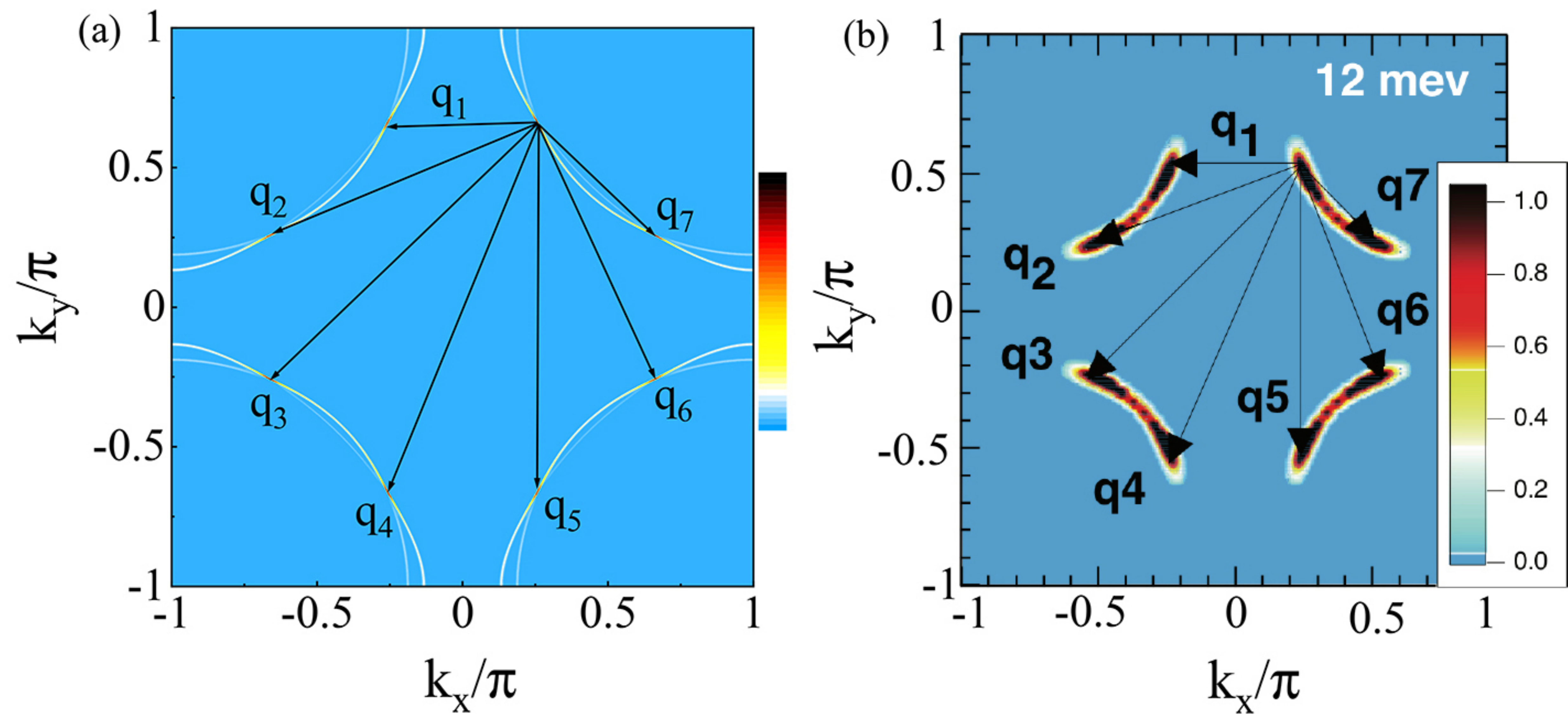}
\caption{(Color online) (a) The intensity map of the quasiparticle excitation spectrum
in the binding-energy $\omega=12$ meV at $\delta=0.15$ with $T=0.002J$ for the impurity
concentration $n_{i}=0.0005$. (b) The corresponding experimental result of the
optimally doped Bi$_{2}$Sr$_{2}$CaCu$_{2}$O$_{8+\delta}$ for $\omega=12$ meV taken from
Ref. \onlinecite{Chatterjee06}. ${\bf q}_{1}$, ${\bf q}_{2}$, ${\bf q}_{3}$, ${\bf q}_{4}$,
${\bf q}_{5}$, ${\bf q}_{6}$, and ${\bf q}_{7}$ indicate different scattering wave vectors.
\label{IS-EFS-MAP}}
\end{figure}

We now turn to discuss the ARPES autocorrelation of cuprate superconductors for a further
understanding of the nature of the quasiparticle excitation. Experimentally, ARPES probes
directly the momentum-space electronic structure of the system
\cite{Damascelli03,Campuzano04,Fink07}, while the ARPES autocorrelation detects directly
the effectively momentum-resolved joint density of states in the electronic state
\cite{Chatterjee06,He14}, yielding the important insights into the nature of the
quasiparticle excitation. On the other hand, scanning tunneling spectroscopy (STS) observes
directly the real-space inhomogeneous electronic structure of the system \cite{Yin21}. In
particular, this STS technique has been also used to infer the momentum-space behavior of
the quasiparticle excitations of cuprate superconductors from the Fourier transform (FT) of
the position- and energy-dependent local density of states (LDOS) $\rho({\bf r},\omega)$,
and then both the real- and momentum-spaces modulations for LDOS are explored simultaneously
\cite{Yin21,Pan01,Hoffman02,Kohsaka07,Kohsaka08,Hamidian16}. The characteristic feature
observed by the FT-STS LDOS $\rho({\bf q},\omega)$ is some sharp peaks at the well-defined
wave vectors ${\bf q}_{i}$ obeying the octet model as shown in Fig. \ref{EFS-MAP},
the quasiparticle scattering interference (QSI)
\cite{Yin21,Pan01,Hoffman02,Kohsaka07,Kohsaka08,Hamidian16} then manifests itself as a
spatial modulation of $\rho({\bf r},\omega)$ with these well-defined wave vector
${\bf q}_{i}$, appearing in the FT-STS LDOS $\rho({\bf q},\omega)$. More importantly, it has
been demonstrated experimentally \cite{Chatterjee06,He14} that the sharp peaks in the ARPES
autocorrelation spectrum are directly correlated with the quasiparticle scattering wave
vectors ${\bf q}_{i}$ connecting the tips of the Fermi arcs in the octet scattering model as
shown in Fig. \ref{EFS-MAP}, and are also well consistent with the QSI peaks observed from
the FT-STS experiments \cite{Yin21,Pan01,Hoffman02,Kohsaka07,Kohsaka08,Hamidian16}. This is
also why the main features of QSI observed in the FT-STS experiments
\cite{Yin21,Pan01,Hoffman02,Kohsaka07,Kohsaka08,Hamidian16} can be also detected from the
ARPES autocorrelation experiments \cite{Chatterjee06,He14}. In this subsection, we further
discuss the influence of the impurity scattering on the electronic state in terms of the
autocorrelation of the quasiparticle excitation spectra.

\begin{figure}[h!]
\centering
\includegraphics[scale=0.17]{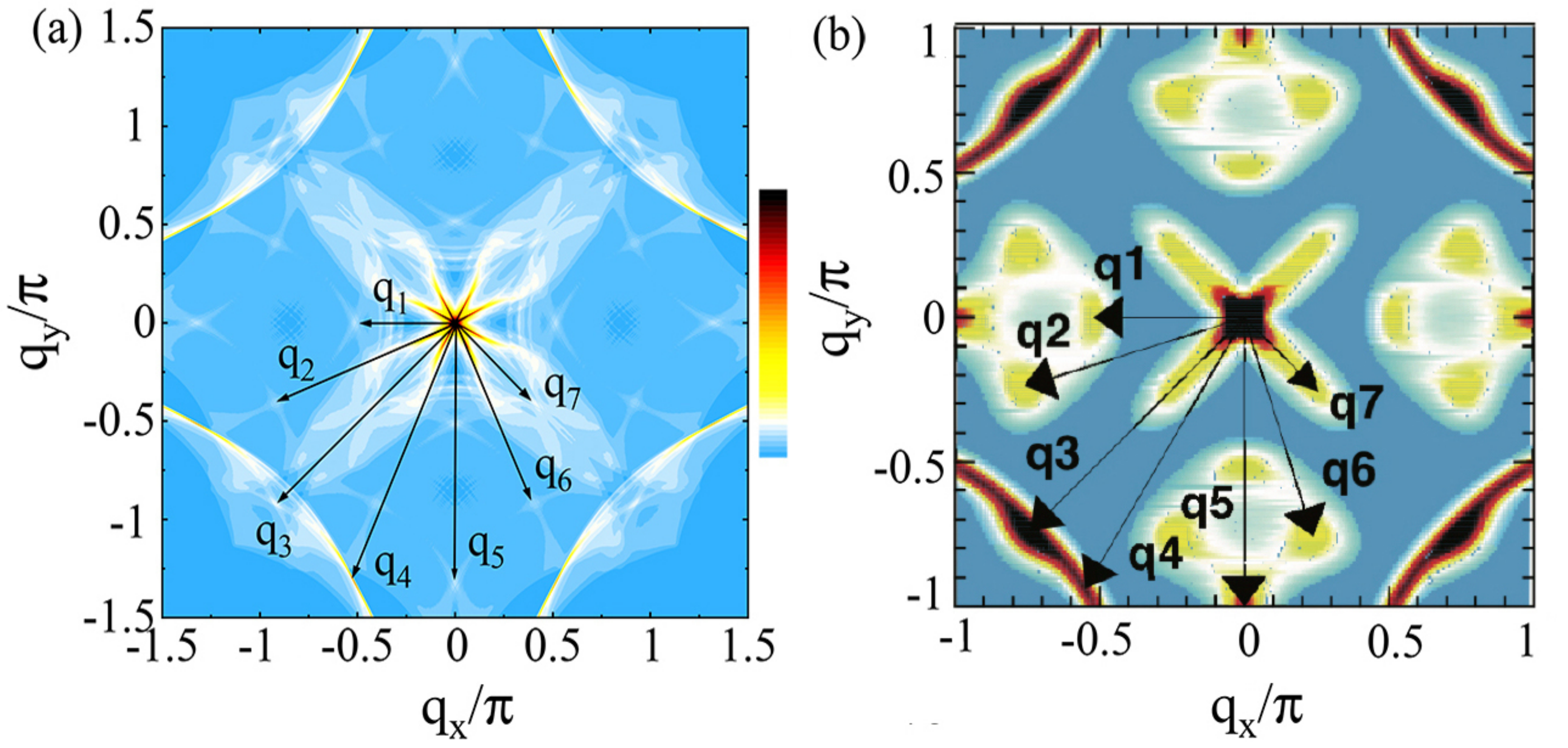}
\caption{(Color online) (a) The intensity map of the autocorrelation spectrum in the
binding-energy $\omega=12$ meV at $\delta=0.15$ with $T=0.002J$ for the impurity
concentration $n_{i}=0.0005$. (b) The corresponding experimental result of the ARPES
autocorrelation spectrum observed from the optimally doped
Bi$_{2}$Sr$_{2}$CaCu$_{2}$O$_{8+\delta}$ for the binding-energy $\omega=12$ meV taken
from Ref. \onlinecite{Chatterjee06}. \label{autocorrelation-maps}}
\end{figure}

The ARPES autocorrelation of cuprate superconductors is described in terms of the
quasiparticle excitation spectrum in Eq. (\ref{IQES}) as \cite{Chatterjee06},
\begin{eqnarray}\label{autocorrelation}
{\bar C}_{\rm I}({\bf q},\omega)={1\over N}\sum_{\bf k}I_{\rm I}({\bf k}+{\bf q},\omega)
I_{\rm I}({\bf k},\omega),
\end{eqnarray}
which measures the autocorrelation of the quasiparticle excitation spectra in
Eq. (\ref{IQES}) at two different momenta ${\bf k}$ and ${\bf k}+{\bf q}$, where the
summation of momentum ${\bf k}$ is restricted within the first BZ just as it has been done
in the experiments \cite{Chatterjee06}. In subsection \ref{Octet-model}, the topology of EFS
(then the zero energy contour) in the pure system has been discussed, where the tips of the
Fermi arcs connected by the scattering wave vectors ${\bf q}_{i}$ construct an {\it octet
scattering model} shown in Fig. \ref{EFS-MAP}. More specifically, this octet scattering
model shown in Fig. \ref{EFS-MAP} can persist into the system in the presence of impurities
at the case for a finite binding-energy \cite{Chatterjee06}. To see this important feature
more clearly, we plot an intensity map of the dressed quasiparticle excitation spectrum
$I_{\rm I}({\bf k},\omega)$ in the case of the binding-energy $\omega=12$ meV at
$\delta=0.15$ with $T=0.002J$ for the impurity concentration $n_{i}=0.0005$ in
Fig. \ref{IS-EFS-MAP}a. For a clear comparison, the corresponding ARPES experimental result
\cite{Chatterjee06} observed on the optimally doped Bi$_{2}$Sr$_{2}$CaCu$_{2}$O$_{8+\delta}$
for the case of the binding-energy $\omega=12$ meV is also shown in Fig. \ref{IS-EFS-MAP}b.
It thus shows that the {\it octet scattering model} with the scattering wave vectors
${\bf q}_{i}$ connecting the tips of the Fermi arcs emerges in the system in the presence of
impurities at the case for a finite binding-energy, which is well consistent with the
corresponding ARPES experimental result \cite{Chatterjee06}.

\begin{figure}[h!]
\centering
\includegraphics[scale=0.18]{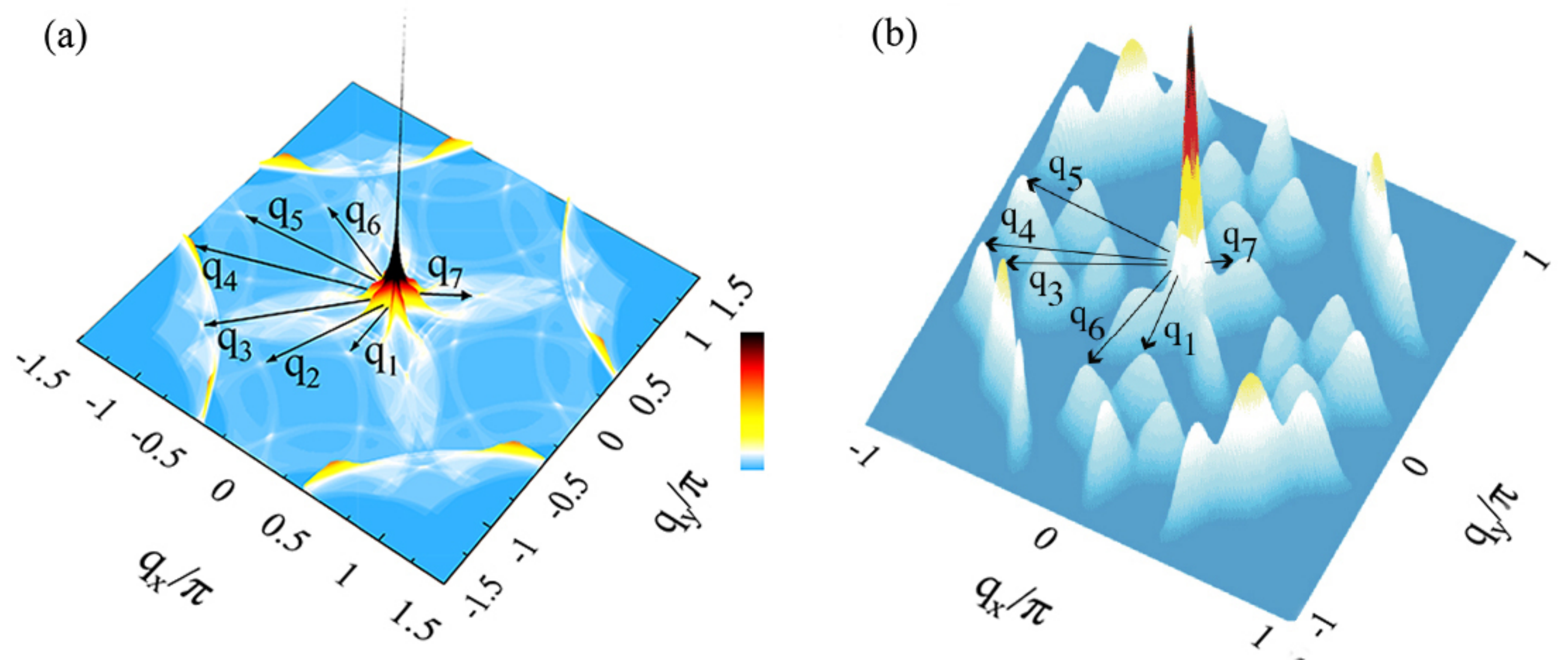}
\caption{(Color online) (a) The surface plot of the autocorrelation spectrum in the
binding-energy $\omega=18$ meV at $\delta=0.15$ with $T=0.002J$ for the impurity
concentration $n_{i}=0.0005$. (b) The corresponding experimental result of the ARPES
autocorrelation observed from the optimally doped Bi$_{2}$Sr$_{2}$CaCu$_{2}$O$_{8+\delta}$
for the binding-energy $\omega=18$ meV taken from Ref. \onlinecite{Chatterjee06}.
\label{autocorrelation-peaks}}
\end{figure}

We are now ready to discuss the ARPES autocorrelation in cuprate superconductors.
In Fig. \ref{autocorrelation-maps}a, we plot the intensity map of the autocorrelation of
the quasiparticle excitation spectra ${\bar C}_{\rm I}({\bf q},\omega)$ in the
binding-energy $\omega=12$ meV at $\delta=0.15$ with $T=0.002J$ for the impurity
concentration $n_{i}=0.0005$. For a better comparison, the corresponding experimental
result \cite{Chatterjee06} detected from the optimally doped
Bi$_{2}$Sr$_{2}$CaCu$_{2}$O$_{8+\delta}$ for the bind-energy $\omega=12$ meV is also
shown in Fig. \ref{autocorrelation-maps}b. Obviously, the corresponding ARPES
experimental result \cite{Chatterjee06} is qualitatively reproduced, where the main
features can be summarized as: (i) there are some discrete spots appear in
${\bar C}_{\rm I}({\bf q},\omega)$, where the joint density of states is highest; (ii)
these discrete spots in ${\bar C}_{\rm I}({\bf q},\omega)$ are directly correlated with
the corresponding wave vectors ${\bf q}_{i}$ connecting the tips of the Fermi arcs in
the octet scattering model shown in Fig. \ref{IS-EFS-MAP}; (iii) the momentum-space
structure of the ARPES autocorrelation pattern of ${\bar C}_{\rm I}({\bf q},\omega)$ is
quite similar to the momentum-space structure of the QSI pattern observed from FT-STS
experiments \cite{Yin21,Pan01,Hoffman02,Kohsaka07,Kohsaka08,Hamidian16}. To see the
autocorrelation pattern of ${\bar C}_{\rm I}({\bf q},\omega)$ more clearly, the surface plot
of ${\bar C}_{\rm I}({\bf q},\omega)$ in the binding-energy $\omega=18$ meV at $\delta=0.15$
with $T=0.002J$ for the impurity concentration $n_{i}=0.0005$ is shown
in Fig. \ref{autocorrelation-peaks}a in comparison with the corresponding experimental result
\cite{Chatterjee06} observed on the optimally doped Bi$_{2}$Sr$_{2}$CaCu$_{2}$O$_{8+\delta}$
for the binding-energy $\omega=18$ meV in Fig. \ref{autocorrelation-peaks}b, where as was
expected, the sharp autocorrelation peaks are located exactly at the discrete spots of
${\bar C}_{\rm I}({\bf q},\omega)$.

In addition to the results plotted in the above Fig. \ref{autocorrelation-maps} and
Fig. \ref{autocorrelation-peaks}, we have also performed a series of calculations for
${\bar C}_{\rm I}({\bf q},\omega)$ with other different sets of the strength of the impurity
scattering at different impurity concentrations as in the case of the discussions in Sec.
\ref{ESEFIS}. Comparing these results together with the results shown in
Fig. \ref{autocorrelation-maps} and Fig. \ref{autocorrelation-peaks} with the corresponding
results in the pure system \cite{Gao19}, we thus find that except for the sharp peaks in the
autocorrelation pattern in the pure system that are broadened by the impurity scattering, (i)
at a given set of the impurity scattering strength, the weight of the extra peaks in the
autocorrelation pattern of the pure system is smoothly depressed when the impurity
concentration level is raised, and (ii) on the other hand, at a given impurity concentration,
the weight of the extra peaks in the autocorrelation pattern of the pure system is gradually
suppressed with the increase of the impurity scattering strength. In other words, the impurity
concentration presents a similar behavior of the impurity scattering strength. More 
importantly, in the reasonable parameter range of the impurity scattering strength and impurity 
concentration, $45J<V_{1}<95J$, $V_{j}<V_{1}$ with $j=2,3,4,...8$, and $0.0004 <n_{i}<0.00055$, 
the weight of the extra peaks in the autocorrelation pattern of the pure system is eliminated 
completely by the impurity scattering as the results shown in Fig. \ref{autocorrelation-maps} 
and Fig. \ref{autocorrelation-peaks}, leading to that the obtained results of the 
autocorrelation pattern as the results shown in Fig. \ref{autocorrelation-maps} and 
Fig. \ref{autocorrelation-peaks} are consistent with the corresponding ARPES experimental 
observations \cite{Chatterjee06} on the optimally doped Bi$_{2}$Sr$_{2}$CaCu$_{2}$O$_{8+\delta}$. 
The qualitative agreement between the theoretical results and experimental observations 
therefore indicate that the unconventional features of the ARPES autocorrelation pattern (then 
the QSI pattern) are dominated by both the strong electron correlation and impurity scattering. 
The present study also shows that the {\it microscopic octet scattering model} obtained based on 
the kinetic-energy-driven superconductivity can give a consistent description of the influence 
of impurities on the electronic structure in cuprate superconductors.

\section{Summary and discussions}\label{conclude}

Starting from the $t$-$J$ model in the fermion-spin representation, we have rederived
the homogenous part of the electron propagator with the d-wave symmetry based on the
kinetic-energy-driven SC mechanism, and shown that the formation of the Fermi arcs is due
to the EFS reconstruction, where a large number of the low-energy electronic states is
available at around the tips of the Fermi arcs, and then the most physical properties of
cuprate superconductors are controlled by the quasiparticle excitations at around the tips
of the Fermi arcs. These tips of the Fermi arcs connected by the scattering wave vectors
${\bf q}_{i}$ naturally construct an {\it octet scattering model}. With the help of this
homogenous electron propagator and the associated octet scattering model, we then have
investigated the influence of the impurity scattering on the electronic structure of cuprate
superconductors within the standard perturbation theory, where although the impurity
scattering is treated in terms of the self-consistent $T$-matrix approach, the impurity
scattering self-energy is evaluated firstly in the {\it Fermi-arc-tip approximation} of the
quasiparticle excitations and scattering processes. The obtained results show that (i) the
quasiparticle band structure is further renormalized by the real part of the impurity
scattering self-energy in the particle-hole channel, while the quasiparticle lifetime is
further reduced by the corresponding imaginary part of the impurity scattering self-energy,
with the renormalization strength and reduction extent that increase as the impurity
concentration is increased; (ii) the impurity scattering self-energy in the particle-particle
channel generates a strong deviation from the d-wave behaviour of the SC gap, where with the
increase of the impurity concentration, the magnitude of the SC gap along EFS is
progressively reduced except for at around the nodal region, where the gap that vanishes in
the pure system opens with the magnitude of the gap that smoothly increases, which therefore
leads to the existence of a finite gap over the entire EFS. Furthermore, we have employed
these impurity scattering self-energies in the particle-hole and particle-particle channels
to study the influence of the impurity scattering on the complicated line-shape in the
quasiparticle excitation spectrum and the ARPES autocorrelation spectrum, and the obtained
results are well consistent with the corresponding experimental observations. Our theory
therefore indicates that the unconventional features of the electronic structure in cuprate
superconductors are generated by both the strong electron correlation and impurity scattering.

The theoretical framework, especially the Fermi-arc-tip approximation, developed in this
paper for the understanding of the influence of the impurity scattering on the electronic
structure of cuprate superconductors can be also employed to study the influence of the
impurity scattering on other various properties of cuprate superconductors both in the
SC- and normal-states. In particular, based on this theoretical framework, we have also
discussed the energy dependence of the SC-state quasiparticle transport in cuprate
superconductors by the consideration of the contributions of the vertex correction
\cite{Zeng22}. These and the related works will be presented elsewhere.

\section*{Acknowledgements}

This work is supported by the National Key Research and Development Program of China under
Grant No. 2021YFA1401803, and the National Natural Science Foundation of China (NSFC) under
Grant Nos. 11974051 and 11734002.

\begin{appendix}

\begin{widetext}

\section{T-matrix equation}
\label{T-matrix-equation}

In this Appendix, we derive explicitly the result of the T-matrix equation
(\ref{T-matrix-6}) of the main text. The self-consistent T-matrix equation
(\ref{T-matrix-5}) can be expanded in the following way,
%\begin{widetext}
\begin{eqnarray}\label{T-iterr-2}
\tilde{T}_{\mu \nu} &=&\bar{V}_{\mu\nu}\otimes\tau_{3} \nonumber\\
&+&\bar{V}_{\mu{\rm A}}\otimes [\tau_{3}\tilde{I}^{\rm (A)}_{\tilde{G}}(\omega)]
\{\bar{V}_{{\rm A}\nu}\otimes\tau_{3}+\bar{V}_{\rm AA}\otimes
[\tau_{3}\tilde{I}^{\rm (A)}_{\tilde{G}}(\omega)](\bar{V}_{{\rm A}\nu}\otimes\tau_{3}
+\bar{V}_{\rm AA}\otimes [\tau_{3}\tilde{I}^{\rm (A)}_{\tilde{G}}(\omega)]
\tilde{T}_{{\rm A}\nu}+\bar{V}_{\rm AB}\otimes
[\tau_{3}\tilde{I}^{\rm (B)}_{\tilde{G}}(\omega)]\tilde{T}_{{\rm B}\nu})\nonumber\\
&+&\bar{V}^{\rm AB}\otimes [\tau_{3}\tilde{I}^{\rm (B)}_{\tilde{G}}(\omega)]
(\bar{V}_{{\rm B}\nu}\otimes\tau_{3}+\bar{V}_{\rm BA}\otimes [\tau_{3}
\tilde{I}^{\rm (A)}_{\tilde{G}}(\omega)]*\tilde{T}_{{\rm A}\nu}+\bar{V}_{\rm BB}\otimes
[\tau_{3}\tilde{I}^{\rm (B)}_{\tilde{G}}(\omega)]*\tilde{T}_{{\rm B}\nu})\} \nonumber\\
&+&\bar{V}_{\mu{\rm B}}\otimes [\tau_{3}\tilde{I}^{\rm (B)}_{\tilde{G}}(\omega)]
\{\bar{V}_{{\rm B}\nu}\otimes\tau_{3}+\bar{V}_{\rm BA}\otimes [\tau_{3}
\tilde{I}^{\rm (A)}_{\tilde{G}}(\omega)](\bar{V}_{{\rm A}\nu}\otimes\tau_{3}+\bar{V}_{\rm AA}
\otimes [\tau_{3}\tilde{I}^{\rm (A)}_{\tilde{G}}(\omega)]\tilde{T}_{{\rm A}\nu}
+\bar{V}_{\rm AB}\otimes [\tau_{3}\tilde{I}^{\rm (B)}_{\tilde{G}}(\omega)]
\tilde{T}_{{\rm B}\nu}) \nonumber\\
&+&\bar{V}_{\rm BB}\otimes [\tau_{3}\tilde{I}^{\rm (B)}_{\tilde{G}}(\omega)]
(\bar{V}_{{\rm B}\nu}\otimes\tau_{3}+\bar{V}_{\rm BA}\otimes [\tau_{3}
\tilde{I}^{\rm (A)}_{\tilde{G}}(\omega)]\tilde{T}_{{\rm A}\nu}+\bar{V}_{\rm BB}
\otimes [\tau_{3}\tilde{I}^{\rm (B)}_{\tilde{G}}(\omega)]\tilde{T}_{{\rm B}\nu})\}
\nonumber\\
&=&\bar{V}_{\mu\nu}\otimes\tau_{3}+\bar{V}_{\mu{\rm A}}\bar{V}_{{\rm A}\nu}\otimes [\tau_{3}
\tilde{I}^{\rm (A)}_{\tilde{G}}(\omega)]\tau_{3}+\bar{V}_{\mu{\rm B}}\bar{V}_{{\rm B}\nu}
\otimes [\tau_{3}\tilde{I}^{\rm (B)}_{\tilde{G}}(\omega)]\tau_{3}
\nonumber\\
&+&\bar{V}_{\mu{\rm A}}\bar{V}_{\rm AA}\bar{V}_{{\rm A}\nu}\otimes [\tau_{3}
\tilde{I}^{\rm (A)}_{\tilde{G}}(\omega)]^{2}\tau_{3}+\bar{V}_{\mu{\rm A}}\bar{V}_{\rm AB}
\bar{V}_{{\rm B}\nu}\otimes [\tau_{3}\tilde{I}^{\rm (A)}_{\tilde{G}}(\omega)][\tau_{3}
\tilde{I}^{\rm (B)}_{\tilde{G}}(\omega)]\tau_{3}
\nonumber\\
&+&\bar{V}_{\mu{\rm B}}\bar{V}_{\rm BA}\bar{V}_{{\rm A}\nu}\otimes [\tau_{3}
\tilde{I}^{\rm (B)}_{\tilde{G}}(\omega)][\tau_{3}\tilde{I}^{\rm (A)}_{\tilde{G}}(\omega)]
\tau_{3}+\bar{V}_{\mu{\rm B}}\bar{V}_{\rm BB}\bar{V}_{{\rm B}\nu}\otimes [\tau_{3}
\tilde{I}^{\rm (B)}_{\tilde{G}}(\omega)]^{2}\tau_{3}
\nonumber\\
&+&\bar{V}_{\mu{\rm A}}\bar{V}_{\rm AA}\bar{V}_{\rm AA}\otimes [\tau_{3}
\tilde{I}^{\rm (A)}_{\tilde{G}}(\omega)]^{3}\tilde{T}_{{\rm A}\nu}+\bar{V}_{\mu{\rm A}}
\bar{V}_{\rm AA}\bar{V}_{\rm AB}\otimes [\tau_{3}\tilde{I}^{\rm (A)}_{\tilde{G}}(\omega)]^{2}
[\tau_{3}\tilde{I}^{\rm (B)}_{\tilde{G}}(\omega)]\tilde{T}_{{\rm B}\nu}
\nonumber\\
&+&\bar{V}_{\mu{\rm A}}\bar{V}_{\rm AB}\bar{V}_{\rm BA}\otimes [\tau_{3}
\tilde{I}^{\rm (A)}_{\tilde{G}}(\omega)][\tau_{3}\tilde{I}^{\rm (B)}_{\tilde{G}}(\omega)]
[\tau_{3}\tilde{I}^{\rm (A)}_{\tilde{G}}(\omega)]\tilde{T}_{{\rm A}\nu}+\bar{V}_{\mu{\rm A}}
\bar{V}_{\rm AB}\bar{V}_{\rm BB}\otimes [\tau_{3}\tilde{I}^{\rm (A)}_{\tilde{G}}(\omega)]
[\tau_{3}\tilde{I}^{\rm (B)}_{\tilde{G}}(\omega)]^{2}\tilde{T}_{{\rm B}\nu}
\nonumber\\
&+&\bar{V}_{\mu{\rm B}}\bar{V}_{\rm BA}\bar{V}_{\rm AA}\otimes [\tau_{3}
\tilde{I}^{\rm (B)}_{\tilde{G}}(\omega)][\tau_{3}\tilde{I}^{\rm (A)}_{\tilde{G}}(\omega)]^{2}
*\tilde{T}_{{\rm A}\nu}+\bar{V}_{\mu{\rm B}}\bar{V}_{\rm BA}\bar{V}_{\rm AB}\otimes [\tau_{3}
\tilde{I}^{\rm (B)}_{\tilde{G}}(\omega)][\tau_{3}\tilde{I}^{\rm (A)}_{\tilde{G}}(\omega)]
[\tau_{3}\tilde{I}^{\rm (B)}_{\tilde{G}}(\omega)]\tilde{T}_{{\rm B}\nu}
\nonumber\\
&+&\bar{V}_{\mu{\rm B}}\bar{V}_{\rm BB}\bar{V}_{\rm BA}\otimes [\tau_{3}
\tilde{I}^{\rm (B)}_{\tilde{G}}(\omega)]^{2}[\tau_{3}\tilde{I}^{\rm (A)}_{\tilde{G}}(\omega)]
\tilde{T}_{{\rm A}\nu}+\bar{V}_{\mu{\rm B}}\bar{V}_{\rm BB}\bar{V}_{\rm BB}\otimes [\tau_{3}
\tilde{I}^{\rm (B)}_{\tilde{G}}(\omega)]^{3}\tilde{T}_{{\rm B}\nu}.
\end{eqnarray}
%\end{widetext}
To solve this self-consistent T-matrix equation, we define a $4\times 4$ unit matrix
$\hat{I}_{\upsilon}$ in the $\bar{V}$-space, and then right multiply the matrix
$\hat{I}_{\upsilon}\otimes\tau_{3}$ in the above T-matrix equation (\ref{T-iterr-2}), which
leads to an iterative T-matrix equation as,
\begin{eqnarray}\label{Tmat-iterat}
\tilde{T}_{\mu\nu}*\hat{I}_{\upsilon}\otimes\tau_{3} &=&
\sum\limits_{\alpha}T^{(\alpha)}_{\mu\nu}\otimes\tau_{\alpha}\tau_{3}
\nonumber\\
&=&\bar{V}_{\mu\nu}\otimes\tau_{0}+\bar{V}_{\mu{\rm A}}\bar{V}_{{\rm A}\nu}\otimes [\tau_{3}
\tilde{I}^{\rm (A)}_{\tilde{G}}(\omega)]+\bar{V}_{\mu{\rm B}}\bar{V}_{{\rm B}\nu}
\otimes [\tau_{3}\tilde{I}^{\rm (B)}_{\tilde{G}}(\omega)]+\bar{V}_{\mu{\rm A}}
\bar{V}_{{\rm AA}}\bar{V}_{{\rm A}\nu}\otimes [\tau_{3}
\tilde{I}^{\rm (A)}_{\tilde{G}}(\omega)]^{2}
\nonumber\\
&+&\bar{V}_{\mu{\rm A}}\bar{V}_{\rm AB}\bar{V}_{{\rm B}\nu}\otimes [\tau_{3}
\tilde{I}^{\rm (A)}_{\tilde{G}}(\omega)][\tau_{3}\tilde{I}^{\rm (B)}_{\tilde{G}}(\omega)]
+\bar{V}_{\mu{\rm B}}\bar{V}_{\rm BA}\bar{V}_{{\rm A}\nu}\otimes [\tau_{3}
\tilde{I}^{\rm (B)}_{\tilde{G}}(\omega)][\tau_{3}\tilde{I}^{\rm (A)}_{\tilde{G}}(\omega)]
+\bar{V}_{\mu{\rm B}}\bar{V}_{\rm BB}\bar{V}_{{\rm B}\nu}\otimes [\tau_{3}
\tilde{I}^{\rm (B)}_{\tilde{G}}(\omega)]^{2}
\nonumber\\
&+&\bar{V}_{\mu{\rm A}}\bar{V}_{\rm AA}\bar{V}_{\rm AA}\otimes [\tau_{3}
\tilde{I}^{\rm (A)}_{\tilde{G}}(\omega)]^{3}\sum\limits_{\alpha}T^{(\alpha)}_{{\rm A}\nu}
\otimes\tau_{\alpha}\tau_{3}+\bar{V}_{\mu{\rm A}}\bar{V}_{\rm AA}\bar{V}_{\rm AB}\otimes
[\tau_{3}\tilde{I}^{\rm (A)}_{\tilde{G}}(\omega)]^{2}[\tau_{3}
\tilde{I}^{\rm (B)}_{\tilde{G}}(\omega)]\sum\limits_{\alpha}T^{(\alpha)}_{{\rm B}\nu}\otimes
\tau_{\alpha}\tau_{3}
\nonumber\\
&+&\bar{V}_{\mu{\rm A}}\bar{V}_{\rm AB}\bar{V}_{\rm BA}\otimes [\tau_{3}
\tilde{I}^{\rm (A)}_{\tilde{G}}(\omega)][\tau_{3}\tilde{I}^{\rm (B)}_{\tilde{G}}(\omega)]
[\tau_{3}\tilde{I}^{\rm (A)}_{\tilde{G}}(\omega)]\sum\limits_{\alpha}T^{(\alpha)}_{{\rm A}\nu}
\otimes\tau_{\alpha}\tau_{3}\nonumber\\
&+&\bar{V}_{\mu{\rm A}}\bar{V}_{\rm AB}\bar{V}_{\rm BB}\otimes [\tau_{3}
\tilde{I}^{\rm (A)}_{\tilde{G}}(\omega)][\tau_{3}\tilde{I}^{\rm (B)}_{\tilde{G}}(\omega)]^{2}
\sum\limits_{\alpha}T^{(\alpha)}_{{\rm B}\nu}\otimes\tau_{\alpha}\tau_{3}\nonumber\\
&+&\bar{V}_{\mu{\rm B}}\bar{V}_{\rm BA}\bar{V}{\rm AA}\otimes [\tau_{3}
\tilde{I}^{\rm (B)}_{\tilde{G}}(\omega)][\tau_{3}\tilde{I}^{\rm (A)}_{\tilde{G}}(\omega)]^{2}
\sum\limits_{\alpha}T^{(\alpha)}_{{\rm A}\nu}\otimes\tau_{\alpha}\tau_{3}\nonumber\\
&+&\bar{V}_{\mu{\rm B}}\bar{V}_{\rm BA}\bar{V}_{\rm AB}\otimes [\tau_{3}
\tilde{I}^{\rm (B)}_{\tilde{G}}(\omega)][\tau_{3}\tilde{I}^{\rm (A)}_{\tilde{G}}(\omega)]
[\tau_{3}\tilde{I}^{\rm (B)}_{\tilde{G}}(\omega)]\sum\limits_{\alpha}T^{(\alpha)}_{{\rm B}\nu}
\otimes\tau_{\alpha}\tau_{3}\nonumber\\
&+&\bar{V}_{\mu{\rm B}}\bar{V}_{\rm BB}\bar{V}_{\rm BA}\otimes [\tau_{3}
\tilde{I}^{\rm (B)}_{\tilde{G}}(\omega)]^{2}[\tau_{3}\tilde{I}^{\rm (A)}_{\tilde{G}}(\omega)]
\sum\limits_{\alpha}T^{(\alpha)}_{{\rm A}\nu}\otimes\tau_{\alpha}\tau_{3}
\nonumber\\
&+&\bar{V}_{\mu{\rm B}}\bar{V}_{\rm BB}\bar{V}_{\rm BB}\otimes [\tau_{3}
\tilde{I}^{\rm (B)}_{\tilde{G}}(\omega)]^{3}\sum\limits_{\alpha}T^{(\alpha)}_{{\rm B}\nu}
\otimes\tau_{\alpha}\tau_{3}.
\end{eqnarray}
Now it is quite easy to verify that the above T-matrix satisfies the following equation,
\begin{eqnarray}\label{T-mat-result}
\sum\limits_{\alpha}T^{(\alpha)}_{\mu\nu}\otimes\tau_{\alpha}\tau_{3}
&=& [\bar{V}\otimes\tau_{0}(1+\tilde{M}+\tilde{M}^{2}+\tilde{M}^{3}+\cdots)]_{\mu\nu}
=\left ( \bar{V}\otimes\tau_{0}{1\over 1-\tilde{M}} \right )_{\mu\nu}
\nonumber\\
&=&\left [ \left (
\begin{array}{cc}
\bar{V}_{\rm AA}\otimes\tau_{0}, & \bar{V}_{\rm AB}\otimes\tau_{0}\\
\bar{V}_{\rm BA}\otimes\tau_{0}, & \bar{V}_{\rm BB}\otimes\tau_{0}
\end{array}\right ){1\over 1- \tilde{M}}\right ]_{\mu\nu},
\end{eqnarray}
where the matrix $\tilde{M}$ has been given in Eq. (\ref{M-matrix}) of the main text.
Following the Einstein summation rule, the inverse matrix $\bar{M} = (1 - \tilde{M})^{-1}$
can be expressed as,
\begin{eqnarray}
\bar{M}=\Lambda^{(\alpha)}\otimes\tau_{\alpha}=
\left(
\begin{array}{ccc}
\Lambda_{\rm AA}^{(\alpha)}\otimes\tau_{\alpha} &\vline & \Lambda_{\rm AB}^{(\alpha)}
\otimes\tau_{\alpha}\\
\hline
\Lambda_{\rm BA}^{(\alpha)}\otimes\tau_{\alpha} &\vline & \Lambda_{\rm BB}^{(\alpha)}
\otimes\tau_{\alpha}
\end{array}\right)
=\left (
\begin{array}{ccccccccc}
\Lambda_{11}^{(\alpha)}\tau_{\alpha} & \Lambda_{12}^{(\alpha)}\tau_{\alpha} &
\cdots  & \Lambda_{14}^{(\alpha)}\tau_{\alpha}& \vline&
\Lambda_{15}^{(\alpha)}\tau_{\alpha}& \Lambda_{16}^{(\alpha)}\tau_{\alpha}&
\cdots  & \Lambda_{18}^{(\alpha)}\tau_{\alpha}\\
\Lambda_{21}^{(\alpha)}\tau_{\alpha} & \Lambda_{22}^{(\alpha)}\tau_{\alpha} &
\cdots  & \Lambda_{24}^{(\alpha)}\tau_{\alpha}& \vline&
\Lambda_{25}^{(\alpha)}\tau_{\alpha}& \Lambda_{26}^{(\alpha)}\tau_{\alpha}&
\cdots & \Lambda_{28}^{(\alpha)}\tau_{\alpha}\\
\vdots & \vdots & \vdots & \vdots &\vline& \vdots & \vdots & \vdots & \vdots \\
\Lambda_{41}^{(\alpha)}\tau_{\alpha} & \Lambda_{42}^{(\alpha)}\tau_{\alpha} &
\cdots & \Lambda_{44}^{(\alpha)}\tau_{\alpha}& \vline&
\Lambda_{45}^{(\alpha)}\tau_{\alpha}& \Lambda_{46}^{(\alpha)}\tau_{\alpha}&
\cdots & \Lambda_{48}^{(\alpha)}\tau_{\alpha}\\
\hline
\Lambda_{51}^{(\alpha)}\tau_{\alpha} & \Lambda_{52}^{(\alpha)}\tau_{\alpha} &
\cdots & \Lambda_{54}^{(\alpha)}\tau_{\alpha}& \vline&
\Lambda_{55}^{(\alpha)}\tau_{\lambda}& \Lambda_{56}^{(\alpha)}\tau_{\alpha}&
\cdots & \Lambda_{58}^{(\alpha)}\tau_{\alpha}\\
\Lambda_{61}^{(\alpha)}\tau_{\alpha} & \Lambda_{62}^{(\alpha)}\tau_{\alpha} &
\cdots & \Lambda_{64}^{(\alpha)}\tau_{\alpha}& \vline&
\Lambda_{65}^{(\alpha)}\tau_{\alpha}& \Lambda_{66}^{(\alpha)}\tau_{\lambda}&
\cdots & \Lambda_{68}^{(\alpha)}\tau_{\lambda}\\
\vdots & \vdots & \vdots & \vdots &\vline& \vdots & \vdots & \vdots & \vdots \\
\Lambda_{81}^{(\alpha)}\tau_{\lambda} & \Lambda_{82}^{(\alpha)}\tau_{\alpha} &
\cdots & \Lambda_{84}^{(\alpha)}\tau_{\alpha}& \vline&
\Lambda_{85}^{(\alpha)}\tau_{\alpha}& \Lambda_{86}^{(\alpha)}\tau_{\alpha}&
\cdots & \Lambda_{88}^{(\alpha)}\tau_{\alpha}\\
\end{array}\right).~~~~
\end{eqnarray}
However, according to the Pauli matrixes $\tau_{0}$, $\tau_1$, $\tau_2$, and $\tau_3$, the
block $[jj']$ of the matrix $\bar{M}$ can be decomposed as,
\begin{eqnarray}
\bar{M}_{jj'}=\sum\limits_{\alpha=0}^{3}\Lambda_{j j'}^{(\alpha)}\tau_{\alpha}
&=&{1\over 2}(\bar{M}_{j j'}+\bar{M}_{j+1 j'+1})\tau_{0}
+{1\over 2}(\bar{M}_{j j'} - \bar{M}_{j+1 j'+1})\tau_{3} \nonumber\\
&+&{1\over 2}(\bar{M}_{j j'+1} +\bar{M}_{j+1 j'})\tau_{1} +
{i\over 2}(\bar{M}_{j j'+1} - \bar{M}_{j+1 j'})\tau_{2},
\end{eqnarray}
and then the final form of the T-matrix in Eq. (\ref{T-mat-result}) can be obtained as,
\begin{eqnarray}\label{T-matrix-15}
\sum\limits_{\alpha}T^{(\alpha)}_{\mu\nu}\otimes\tau_{\alpha}\tau_{3}
&=& \left [ \left (
\begin{array}{ccc}
\bar{V}_{\rm AA}\otimes\tau_{0} & \vline & \bar{V}_{\rm AB}\otimes\tau_{0}\\
\hline
\bar{V}_{\rm BA}\otimes\tau_{0} & \vline & \bar{V}_{\rm BB}\otimes\tau_{0}
\end{array}
\right )
\left(
\begin{array}{ccc}
\Lambda_{\rm AA}^{(\alpha)}\otimes\tau_{\alpha} &\vline & \Lambda_{\rm AB}^{(\alpha)}
\otimes\tau_{\alpha}\\
\hline
\Lambda_{\rm BA}^{(\alpha)}\otimes\tau_{\alpha} &\vline & \Lambda_{\rm BB}^{(\alpha)}
\otimes\tau_{\alpha}
\end{array}
\right)\right ]_{\mu\nu}
\nonumber\\
&=& \sum\limits_{\mu'= {\rm A}}^{\rm B}\sum\limits_{\alpha=0}^{3}\bar{V}_{\mu\mu'}
\Lambda^{(\alpha)}_{\mu'\nu}\otimes\tau_{\alpha}
=\sum\limits_{\alpha=0}^{3}(\sum\limits_{\mu'={\rm A}}^{\rm B}\bar{V}_{\mu\mu'}
\Lambda^{(\alpha)}_{\mu'\nu})\otimes\tau_{\alpha},
\end{eqnarray}
which is the same as quoted in Eq. (\ref{T-matrix-6}) of the main text.

\end{widetext}

\end{appendix}

\end{document}